\input epsf
\documentstyle{amsppt}
\overfullrule=0pt
\newcount\mgnf\newcount\tipi\newcount\tipoformule\newcount\greco
\tipi=2          %uso caratteri: 2=cmcompleti, 1=cmparziali, 0=amparziali
\tipoformule=0   %=0 da numeroparagrafo.numeroformula; se no numero
                 %assoluto

\global\newcount\numsec\global\newcount\numfor
\global\newcount\numapp\global\newcount\numcap
\global\newcount\numfig\global\newcount\numpag
\global\newcount\numnf
\global\newcount\numtheo

\def\SIA #1,#2,#3 {\senondefinito{#1#2}%
\expandafter\xdef\csname #1#2\endcsname{#3}\else
\write16{???? ma #1,#2 e' gia' stato definito !!!!} \fi}

\def \FU(#1)#2{\SIA fu,#1,#2 }

\def\etichetta(#1){(\veroparagrafo.\veraformula)%
\SIA e,#1,(\veroparagrafo.\veraformula) %
\global\advance\numfor by 1%
\write15{\string\FU (#1){\equ(#1)}}%
\write16{ EQ #1 ==> \equ(#1)  }}

\def\etichettat(#1){\veroparagrafo.\veratheorema:%
\SIA e,#1,{\veroparagrafo.\veratheorema} %
\global\advance\numtheo by 1%
\write15{\string\FU (#1){\thu(#1)}}%
\write16{ TH #1 ==> \thu(#1)  }}

\def\etichettaa(#1){(A\veraappendice.\veraformula)
 \SIA e,#1,(A\veraappendice.\veraformula)
 \global\advance\numfor by 1
 \write15{\string\FU (#1){\equ(#1)}}
 \write16{ EQ #1 ==> \equ(#1) }}
\def\getichetta(#1){Fig. \verafigura
 \SIA g,#1,{\verafigura}
 \global\advance\numfig by 1
 \write15{\string\FU (#1){\graf(#1)}}
 \write16{ Fig. #1 ==> \graf(#1) }}
\def\retichetta(#1){\numpag=\pgn\SIA r,#1,{\verapagina}
 \write15{\string\FU (#1){\rif(#1)}}
 \write16{\rif(#1) ha simbolo  #1  }}
\def\etichettan(#1){(n\verocapitolo.\veranformula)
 \SIA e,#1,(n\verocapitolo.\veranformula)
 \global\advance\numnf by 1
\write16{\equ(#1) <= #1  }}

\newdimen\gwidth
\gdef\profonditastruttura{\dp\strutbox}
\def\senondefinito#1{\expandafter\ifx\csname#1\endcsname\relax}
\def\BOZZA{
\def\alato(##1){
 {\vtop to \profonditastruttura{\baselineskip
 \profonditastruttura\vss
 \rlap{\kern-\hsize\kern-1.2truecm{$\scriptstyle##1$}}}}}
\def\galato(##1){ \gwidth=\hsize \divide\gwidth by 2
 {\vtop to \profonditastruttura{\baselineskip
 \profonditastruttura\vss
 \rlap{\kern-\gwidth\kern-1.2truecm{$\scriptstyle##1$}}}}}
\def\verapagina{
{\romannumeral\number\numcap}.\number\numsec.\number\numpag}}

\def\alato(#1){}
\def\galato(#1){}
\def\veroparagrafo{\number\numsec}\def\veraformula{\number\numfor}
\def\veraappendice{\number\numapp}
\def\verapagina{\number\pageno}\def\veranformula{\number\numnf}
\def\verafigura{{\romannumeral\number\numcap}.\number\numfig}
\def\verocapitolo{\number\numcap}\def\veranformula{\number\numnf}
\def\veratheorema{\number\numtheo}
\def\Eqn(#1){\eqno{\etichettan(#1)\alato(#1)}}
\def\eqn(#1){\etichettan(#1)\alato(#1)}
\def\TH(#1){{\etichettat(#1)\alato(#1)}}%\csname fu#1\endcsname\fi}
\def\thv(#1){\senondefinito{fu#1}$\clubsuit$#1\else\csname fu#1\endcsname\fi}
\def\thu(#1){\senondefinito{e#1}\thv(#1)\else\csname e#1\endcsname\fi}
\def\ver{\veroparagrafo}
\def\Eq(#1){\eqno{\etichetta(#1)\alato(#1)}}
\def\eq(#1){\etichetta(#1)\alato(#1)}
\def\Eqa(#1){\eqno{\etichettaa(#1)\alato(#1)}}
\def\eqa(#1){\etichettaa(#1)\alato(#1)}
\def\dgraf(#1){\getichetta(#1)\galato(#1)}
\def\drif(#1){\retichetta(#1)}

\def\eqv(#1){\senondefinito{fu#1}$\clubsuit$#1\else\csname fu#1\endcsname\fi}
\def\equ(#1){\senondefinito{e#1}\eqv(#1)\else\csname e#1\endcsname\fi}
\def\graf(#1){\senondefinito{g#1}\eqv(#1)\else\csname g#1\endcsname\fi}
\def\rif(#1){\senondefinito{r#1}\eqv(#1)\else\csname r#1\endcsname\fi}
%%%%%%%%%%%%%%%%%%%%%%%%%%%%%%%%%%%%%%%%%%%%%%%%%%%%%%%%%%%%%%
%%%%%%%%%%%%%%%%%% Numerazione verso il futuro ed eventuali paragrafi
%%%%%%%%%%%%%%%%%% precedenti non inseriti nella scheda da compilare
%%%%%%%%%%%%%%%%%% e elenco referenze bibliografiche creato in
%%%%%%%%%%%%%%%%%% \jobname.bib
\def\bib[#1]{[#1]\numpag=\pgn
\write13{\string[#1],\verapagina}}

\def\include#1{
\openin13=#1.aux \ifeof13 \relax \else
\input #1.aux \closein13 \fi}

\openin14=\jobname.aux \ifeof14 \relax \else
\input \jobname.aux \closein14 \fi
\openout15=\jobname.aux%\write15
\openout13=\jobname.bib
%%%%%%%%%%%%%%%%%%%%%%%%%%%%

%%%%%%%%%%%%%%%%%%%%%%%%%%%%%%%%%%%%%%%%%%%%%%%%%%%%%%%%%%%%%%

\ifnum\tipoformule=1\let\Eq=\eqno\def\eq{}\let\Eqa=\eqno\def\eqa{}
\def\equ{}\fi

%%%%%%%%%%%%%%%%%%%%%%%%%%%%%%%%%%%%%%%%%%%%%%
%%%%%%%%%%%%%%%%%%%%%  Numerazione pagine

{\count255=\time\divide\count255 by 60 \xdef\hourmin{\number\count255}
        \multiply\count255 by-60\advance\count255 by\time
   \xdef\hourmin{\hourmin:\ifnum\count255<10 0\fi\the\count255}}

\def\oramin{\hourmin }

\def\data{\number\day/\ifcase\month\or january \or february \or march \or
april \or may \or june \or july \or august \or september
\or october \or november \or december \fi/\number\year;\ \oramin}

\def\titdate{ \ifcase\month\or January \or February \or March \or
April \or May \or June \or July \or August \or September
\or October \or November \or December \fi \number\day, \number\year;\ \oramin}
% Date to be used for title page

\setbox200\hbox{$\scriptscriptstyle \data $}

\newcount\pgn \pgn=1
\def\foglio{\number\numsec:\number\pgn
\global\advance\pgn by 1}
\def\foglioa{A\number\numsec:\number\pgn
\global\advance\pgn by 1}

\footline={\rlap{\hbox{\copy200}}\hss\tenrm\folio\hss}
%\footline={\hss\tenrm\folio\hss}

%%%%%%%%

\global\newcount\numpunt

\magnification=\magstephalf
%\magnification=1200
\baselineskip=16pt
\parskip=8pt

\voffset=2.5truepc
\hoffset=0.5truepc
\hsize=6.1truein
\vsize=8.4truein %(including running head)
%\vsize=17truecm
%\hsize=11.5truecm
{\headline={\ifodd\pageno\rightheadline \else \leftheadline \fi}}
\def\rightheadline{\it  {tralala}\hfil\tenrm\folio}
\def\leftheadline{\tenrm \folio \hfil\it  {Section $\ver$}}

\def\a{\alpha}
\def\b{\beta}
\def\d{\delta}
\def\e{\epsilon}

\def\f{\phi}
\def\g{\gamma}

\def\s{\sigma}

\def\L{\Lambda}

\def\S{\Sigma}

\def\1{{1\kern-.25em\roman{I}}}
\def\eu{{1\kern-.25em\roman{I}}}
\def\f1{{1\kern-.25em\roman{I}}}

\def\R{{\Bbb R}}  %%
\def\P{{\Bbb P}}  %% carateri piu belle per campi di nombre
\def\E{{\Bbb E}}  %%
  %%

  %%

%\def\P{\hskip.2em\hbox{\rm P\kern-0.8em{I}\hskip.7em}}

% Spezielle Definitionen

\let\cal=\Cal
\def\AA{{\cal A}}

\def\CC{{\cal C}}
\def\DD{{\cal D}}

\def\PP{{\cal P}}

\def\RR{{\cal R}}

\def\chap #1#2{\line{\ch #1\hfill}\numsec=#2\numfor=1\numtheo=1}

\def\wt{\widetilde}

%   Non-character macros

%\newcount\foot
%\foot=1
%\def\note#1{\footnote{${}^{\number\foot}$}{\ftn #1}\advance\foot by 1}
%\def\note#1{\plainfootnote{${}^{\number\foot}$}{\ftn #1}\advance\foot\by 1}
\def\note#1{\footnote{#1}}

\def\frac#1#2{{#1\over #2}}

\def\text#1{\quad{\hbox{#1}}\quad}
\def\newpage{\vfill\eject}
\def\proposition #1{\noindent{\thbf Proposition #1}}

\def\theo #1{\noindent{\thbf Theorem {#1} }}

\def\lemma #1{\noindent{\thbf Lemma {#1} }}

\def\endproof{$\diamondsuit$}
\def\remark{\noindent{\bf Remark: }}
\def\thanks{\noindent{\bf Acknowledgements: }}

\def\var{\hbox{var}}

%\font\thbf=cmcsc10 scaled\magstep1
\font\thbf=cmbxsl10 scaled\magstephalf
% Font-Definitions

\font\ch=cmbx12
\font\ftn=cmr8

\font\it=cmti10
\font\bf=cmbx10

\newfam\msafam
\newfam\msbfam
\newfam\eufmfam
%
% -------------------------------------------------- math macros --------
%
%
\def\hexnumber#1{%
  \ifcase#1 0\or 1\or 2\or 3\or 4\or 5\or 6\or 7\or 8\or
  9\or A\or B\or C\or D\or E\or F\fi}
\font\tenmsa=msam10
\font\sevenmsa=msam7
\font\fivemsa=msam5
\textfont\msafam=\tenmsa
\scriptfont\msafam=\sevenmsa
\scriptscriptfont\msafam=\fivemsa
\edef\msafamhexnumber{\hexnumber\msafam}%
%
%   \mathchardef\restriction"1\msafamhexnumber16
%   "class, family, position (found on amstex guide)
%
\mathchardef\restriction"1\msafamhexnumber16
\mathchardef\ssim"0218
\mathchardef\square"0\msafamhexnumber03
\mathchardef\eqd"3\msafamhexnumber2C
\def\QED{\ifhmode\unskip\nobreak\fi\quad
  \ifmmode\square\else$\square$\fi}
\font\tenmsb=msbm10
\font\sevenmsb=msbm7
\font\fivemsb=msbm5
\textfont\msbfam=\tenmsb
\scriptfont\msbfam=\sevenmsb
\scriptscriptfont\msbfam=\fivemsb
\def\Bbb#1{\fam\msbfam\relax#1}
\font\teneufm=eufm10
\font\seveneufm=eufm7
\font\fiveeufm=eufm5
\textfont\eufmfam=\teneufm
\scriptfont\eufmfam=\seveneufm
\scriptscriptfont\eufmfam=\fiveeufm

\def\({\left(}
\def\){\right)}
%
%-------------------------------------------------------------------
%
% ------- Per compatibilita'
%

%
%\numsec=1\numfor=1

{\headline={\ifodd\pageno\rightheadline \else \leftheadline \fi}}
\def\rightheadline{\it  {CREM and CSB}\hfil\tenrm\folio}
\def\leftheadline{\tenrm \folio \hfil\it  {Section $\ver$}}

\font\tit=cmbx12
\font\aut=cmbx12
\font\aff=cmsl12
\overfullrule=0pt
\def\s{\char'31}
%\nopagenumbers
%{$  $}
%\vskip1.5truecm
\centerline{\tit Poisson convergence in the
 restricted $k$-partioning problem  \note{Work supported in part by the
DFG research center matheon and the 
 European Science Foundation in the programme RDSES}}   
\vskip.1truecm
\centerline{\tit }
\vskip.2truecm
%\vskip.2truecm
%\centerline{\tit }
\vskip1truecm
%\centerline{\tit }
%\vskip2.5cm
%\input epsf
\centerline{\aut Anton Bovier
\note{ e-mail:
bovier\@wias-berlin.de} 
}
\vskip.1truecm
\centerline{\aff Weierstra\s {}--Institut}
\centerline{\aff f\"ur Angewandte Analysis und Stochastik}
\centerline{\aff Mohrenstrasse 39, D-10117 Berlin, Germany}
\centerline{and}
\centerline{\aff Institut f\"ur Mathematik}
\centerline{\aff Technische Universit\"at Berlin}
\centerline{\aff Strasse des 17. Juni 136}
\centerline{\aff 12623 Berlin, Germany}
\vskip0.4truecm
\centerline{\aut  Irina Kurkova\note{\ftn
e-mail: kourkova\@ccr.jussieu.fr}}
\vskip.1truecm
\centerline{\aff Laboratoire de Probabilit\'es et Mod\`eles
Al\'eatoires}
\centerline{\aff Universit\'e Paris 6}
\centerline{\aff 4, place Jussieu, B.C. 188}
\centerline{\aff 75252 Paris, Cedex 5, France}

\vskip0.5truecm\rm
\def\s{\sigma}
\noindent {\bf Abstract:} The randomized $k$-number partitioning
problem is the task to distribute $N$ i.i.d. random variables into $k$
groups in such a way that the sums of the variables in each group are
as similar as possible. The restricted $k$-partitioning problem
refers to the case where the number of elements in each group is fixed
to $N/k$.  In the case $k=2$ it has been shown that the properly
rescaled differences of the two sums in the close to optimal
partitions converge to a Poisson point process, as if they were
independent random variables. We generalize this result to the case
$k>2$ in the restricted problem and show that the vector of differences
between the $k$ sums converges to a $k-1$-dimensional Poisson point process.


%\end

\newpage
{\headline={\ifodd\pageno\rightheadline \else \leftheadline \fi}}
\def\rightheadline{\it  {Number partitioning}\hfil\tenrm\folio}
\def\leftheadline{\tenrm \folio \hfil\it  {Section $\ver$}}

\chap{1. Introduction.}1

The number partitioning problem is a classical problem from
combinatorial optimization. One considers $N$ numbers
$x_1,\dots,x_N$ and one seeks to partition the set $\{1,\dots,
N\}$ into $k$ disjoint subsets $I_1,\dots, I_k$, such that the
sums $K_\b\equiv K_\b(I_1,\dots,I_k) \equiv\sum_{n\in I_\b} x_n$
 are as similar to  each
other as possible. This problem can be cast into the language of
mean field spin systems \cite{Mer1,Mer2,BFM} by realizing that the
set of partitions is equivalent to the set of Potts spin variables
$\s:\{1,\dots,N\}\rightarrow\{1,\dots,k\}^N$. We then define  the
variables
$$
K_\b(\s)\equiv \sum_{n=1}^N x_n\1_{\s_n=\b}, \ \ \b=1,\ldots, k.
\Eq(1.1)
$$
One may introduce a ``Hamiltonian'' as \cite{Mer1,BFM}
$$
H_N(\s) \equiv \sum_{\b=1}^{k-1} |K_{\b}(\s)-K_{\b+1}(\s)|
\Eq(1.2)
$$
and study the minimization problem of this Hamiltonian. In particular, if the
numbers $x_i$ are considered as
random variables, the problem transforms into the
study of a random mean field spin model.
For a detailed discussion we refer to the recent paper \cite{BFM}.

Mertens \cite{Mer1,Mer2} has argued that the problem is close to
the so-called {\it Random Energy Model (REM)}, i.e.\ that the
random variables $K_\b(\s)$ can effectively be considered as
independent random variables for different realizations of $\s$,
at least as far as their extremal properties are concerned. This
claim was proven rigorously in a paper by Borgs et al.\cite{BCP}
in the
 case $k=2$ (see also \cite{BCMP}).

In this paper we extend this result to the
case of arbitrary $k$ and under the additional constraint that
the cardinalities of the sets $I_j$ are all equal.
We formulate this result in the language of
multi-dimensional extremal process.

Let $X_1,\ldots, X_N$ be independent uniformly distributed on [0,1]
   random variables.
   (We assume that $N$ is  always a multiple of $k$.)
     Consider   the state space
 of  configurations $\s$ of $N$ spins,
 where each spin takes $k$ possible values
 $\s=(\s_1,\ldots, \s_N)\in \{1,\ldots, k\}^N$.
      We will restrict ourselves to  configurations
  such that the number of spins taking each value
 equals $N/k$, i.e.\ $\# \{n: \s_n=\b\}=N/k$ for all
    $\b=1,\ldots, k$. Finally, we must
   take  equivalence classes of these configurations: each class
  includes $k!$ configurations
obtained by a permutation of the values of spins  $1,\ldots, k$.
 We denote by  $\Sigma_N$  the state space of these equivalence classes.
     Then
$$
 |\Sigma_N|= {N \choose N/k} {N(1-1/k) \choose N/k} \cdots
  {2N/k \choose N/k} (k!)^{-1} \sim k^N  (2\pi N)^{\frac{1-k}{2}}
k^{\frac{k}{2}} (k!)^{-1}\equiv S(k,N).
\Eq(sin)
$$
 Each configuration $\sigma \in \Sigma_N$ corresponds to a
 partition
   of  $X_{1}, \ldots, X_N$ into $k$ subsets of $N/k$ random
   variables,
  each subset being $\{X_n: \s_n=\b\}$, $\b=1,\ldots, k$.
Then the vector $\vec Y(\s)=\{ Y^{\b}(\s)\}_{\b=1}^{k-1}$ with the
  coordinates
$$
 Y^{\b}(\s)=K_{\b}(\s)-K_{\b+1}(\s)=\sum_{n=1}^N X_n (\1_{\{\s_n=\b\}}-
\1_{\{\s_n=\b+1\}}), \ \ \b=1,\ldots, k-1,
\Eq(az)
$$
 measures the differences of the sums over the subsets.
   Our objective is
   to minimize its norm
    as most as possible.
Our main result is the following theorem.

\theo{\TH(1.1th)}{\it Let
$$
V^{\b}(\s) =k^{\frac{N}{k-1}}  (2\pi N)^{-1}
   k^{\frac{2k-1}{2k-2}}
(k!)^{\frac{-1}{k-1}}2\sqrt{6}|Y^{\b}(\s)|, \ \ \ \ \b=1,\ldots, k-1.
\Eq(v.1)
$$
  Then the point process on $\R_+^{k-1}$
$$
\sum_{\s \in \S_N}
    \delta_{(V^{1} (\s),\ldots, V^{k-1}(\s))}
$$
converges weakly to the Poisson point process on $\R_+^{k-1}$
    with intensity measure given by the  Lebesgue measure.
}

Clearly, from this result we can deduce extremal properties of
$H_N(\s) =\sum_{\b=1}^{k-1} |Y^\b(\s)|
$ straightforwardly.

\noindent{\bf Remark: Integer partitioning problem.}
       It is very easy to derive also  from our Theorem~\thv(1.1th)
   the analogous result for the integer partitioning problem.
        Let $S_1,\ldots, S_N$ be discrete random variables
   uniformly distributed on $\{1,2,\ldots,M(N)\}$ where $M(N)>1$
   is an integer number depending on $N$.
  Let us define
   $$ D^\b(\s)=\sum_{n=1}^N S_n(\1_{\{\s_n=\b\}}-  \1_{\{\s_n=\b+1\}} ).$$

\theo{\TH(1.1bis)}{\it Assume that $M(N)\to \infty$
   as $N\to \infty$ such that $\lim_{N\to \infty}(M(N))^{-1} k^{N/(k-1)}=0$. Let
$$
W^{\b}(\s) = M(N)^{-1} k^{\frac{N}{k-1}}  (2\pi N)^{-1}
   k^{\frac{2k-1}{2k-2}}
(k!)^{\frac{-1}{k-1}}2\sqrt{6}|D^{\b}(\s)|, \ \ \ \ \b=1,\ldots, k-1.
\Eq(v.2)
$$
  Then the point process on $\R_+^{k-1}$
$$
\sum_{\s \in \S_N}
    \delta_{(W^{1} (\s),\ldots, W^{k-1}(\s))}
$$
converges weakly to the Poisson point process on $\R_+^{k-1}$
    with the intensity measure which is the Lebesgue measure.
}

\noindent{\it Proof.} It follows from Theorem \thv(1.1th) by the same
   coupling argument as in the proof of Theorem~6.4 of \cite{BCP}.

The difficulty one is confronted with when
 proving Theorem \thv(1.1th) is that the
standard criteria for convergence for extremal processes to Poisson processes
that go beyond the i.i.d.\ case either assume  independence, stationarity, and
some mixing conditions (see \cite{LLR}), or exchangeability and a very strong
form of asymptotic independence of the finite dimensional marginals
\cite{Gal,BM}.
In the situation at hand, we certainly do not have independence,
or stationarity, nor do we have exchangeability. Worse, also the
asymptotic factorization of marginals does not hold uniformly in the form
 required e.g. in \cite{BM}.

What saves the day is, however, that the asymptotic factorization
conditions hold {\it on average} on $\Sigma_N$,
  and that one can prove a general criterion for
Poisson convergence that requires just that.

Thus the proof of Theorem \thv(1.1th) involves two steps.
In Section 2 we prove an
abstract theorem that gives a
criteria for the convergence of an extremal process
to a Poisson process, and in Sections~3,4
  we show that these are satisfied in the
problem at hand.

   Unfortunately, and this makes the proof seriously
tedious,  for certain vectors $\s,\s'$, there appear very strong
correlations between
  $\vec Y(\s)$ and
   $\vec Y(\s')$
 that have to be dealt with. Such a problem did already
appear in a milder form in the work of Borgs et al \cite{BCP}
  for $k=2$, but in
the general case $k>2$
 the associated linear algebra problems get much
more difficult.

\remark {\bf The unrestricted problem.}
     These linear algebra problems prevented  us to complete the study
 of the unrestricted problem
 (that is when the sets $I_1,\ldots, I_k$
  are not necessarily of size $N/k$) in the case $k>2$.
    In Section~5 we give a conjecture for the
 result similar  to Theorem \thv(1.1th)
 in this case and explain the drawback in the proof
 that remains  to be filled in.

\remark{\bf Dynamical search algorithms.}
It would be interesting to investigate rigorously the properties of dynamical
search algorithms, resp.\ Glauber dynamics associated to this model.
This problem has been studied mainly numerically in a recent paper by
Junier and Kurchan \cite{JK}. They argued that the dynamics for long times
should be described by an effective trap model, just as in the case of the
Random Energy Model. This is clearly going to be the case if the
particular updating rules used in \cite{BBG1}, \cite{BBG2}
 for the REM will be employed,
namely if the transition probability $p(\s,\s')$
depends only on the energy of the initial configuration.
In the REM this choice could be partly justified by the observation that
the deep traps had energies of the order $-N$, while all of their neighbors,
typically, would have energies of the order of 1, give or take $\sqrt{\ln N}$.
Thus, whatever the choice of the dynamics, the main obstacle to motion
will always be the first step away from a deep well.

In the number partitioning problem, the situation is quite different.
Let us only consider the case $k=2$. If $\s$ is one of the very deep
wells, then
$$
H_N(\s)=|\sum_{i=1}^N x_i\s_i|\approx 2^{-N}\sqrt N.
\Eq(a.1)
$$
If $\s^j$ denotes the configuration obtained from $\s$ by inverting one
spin, then
$$
H_N(\s^j)\sim 2|x_j|.
\Eq(a.2)
$$
For a typical sample of $x_i$'s, these values range from
$O(1/N)$ to $1-O(1/N)$. Thus, if we use e.g.\ the Metropolis updating rule,
then the probability of a step from $\s$ to $\s^j$ will be
$\sim \exp(-2\b |x_j|)$. It is by no means clear how high  the saddle point
between two deep wells will be, and whether they will all be of the same order.
This implies
  that the actual time scale for transition times  between deep wells
is not obvious, nor it is clear what the trap model describing the
long term dynamics would have to be.

Of course, changing the Hamiltonian from $ H(\s) $ to $\ln H(\s)$,
as was proposed in \cite{JK}, changes the foregoing
discussion completely and brings us back to the more REM-like situation.

\thanks We thank Stephan Mertens for introducing us to the number
partitioning problem and for valuable discussions.
\bigskip
\chap{2. A general extreme value theorem.}2

Consider series  of $M$  random vectors  $\vec
  V_{i,M}=(V_{i,M}^{1},\ldots, V_{i,M}^p)\in \R_+^p$,
  $i=1,\ldots, M$.

\noindent{\it Notation.}
     We write $\sum_{\a(l)}$ when the sum is taken
   over all possible {\it ordered\/} sequences of {\it different\/}
   indices
   $\{i_1,\ldots, i_l\} \subset \{1,\ldots, M\}$.
   We  also write $\sum_{\a(r_1),\ldots, \a(r_R)}(\cdot)$
  when the sum is taken over  all possible
  {\it ordered\/} sequences of disjoint  ordered subsets
  $\a(r_1)=(i_1,\ldots, i_{r_1})$,
  $\a(r_2)=(i_{r_1+1},\ldots, i_{r_2}),\ldots,$
    $\a(r_R)=(i_{r_1+\cdots +r_{R-1}+1},\ldots, i_{r_1+\cdots+ r_{R}})$
   of $\{1,\ldots, M\}$.

\theo{\TH(mainth)}{\it Assume that for all finite $l=1,2,\ldots $
 and all set of  constants  $c_{j}^{\b}>0$,  $j=1,\ldots, l$,
  $\beta=1,\ldots, p$ we have
$$
\sum_{\a(l)=(i_1,\ldots,i_l)}\P
  \Big(V_{i_j, M}^{\beta} <c_j^{\beta}\  \forall j=1,\ldots, l,
    \b=1,\ldots, p\Big)\to \prod_{j=1,\ldots, l\atop
  \beta =1,\ldots, p}c_{j}^\b, \ \ M\to \infty.
\Eq(maincond)
$$
Then the point process
$$
\Pi_M^p=\sum_{i=1}^M \delta_{(
V_{i,M}^{1},\ldots, V_{i,M}^p)}
\Eq(pp)
$$
  on $\R_+^p$
converges weakly as $M\to \infty$  to the Poisson point process
$\PP^p$ on $\R_+^{p}$
   with the intensity measure which is the Lebesgue measure.}

\noindent{\it Proof.}
  Denote by $\Pi_M^p(A)$ the number of points of
the process
$\Pi_M^p$ in a subset $A\subset \R_+^{p}$.

   The proof of this theorem follows from Kallenberg theorem \cite{Kal}
  on the week convergence of a point process $\Pi_M^p$ to the Poisson process
  $\Pi^p$.
   Applying his theorem in our situation weak convergence holds
  whenever

\noindent(i) For all cubes $A=\prod_{\beta=1}^p[a^{\beta}, b^{\beta})$
 $$
\E \Pi_M^p(A)
   \to |A|,\ \ \  M\to \infty.
\Eq(AB.1000)
$$

\noindent(ii) For all finite union $A=\bigcup_{l=1}^{L}
  \prod_{\beta=1}^p[a^{\beta}_l, b^{\beta}_l)$ of disjoint cubes
$$
\P(\Pi_M^p(A)=0)\to \e^{-|A|}, \ \ \  M\to \infty.
\Eq(AB.1001)
$$

     Our main tool of checking (i) and (ii) is the inclusion-exclusion
  principle which can be summarized as follows: for any $l=1,2,\ldots$
   and any events $O_1,\ldots,O_l$
$$
 \P\Big(\bigcap_{ i=1,\ldots,l}O_i\Big)
   =\sum_{k=0}^{l} \sum_{\AA_k=(i_1,\ldots, i_k)\subset \{1,\ldots, l\}
    \atop i_1<i_2<\cdots<i_k}
    {(-1)^k}\P\Big( \bigcap_{j=1}^k \bar O_{i_j}\Big)
   \Eq(inex)
$$
   where $\bar O_{i_j}$ are complementary events to $O_{i_j}$.
    We use \eqv(inex)
    to  ``invert'' the inequalities
  of type $\{V_{i,M}^\b\geq a^{\b}\}$, i.e.\ to represent
  their probability as the sum of probabilities of
  opposite events, that can be estimated by \eqv(maincond).
   The power of the inclusion-exclusion principle comes from the
  fact that
    the partial sums of the
    right-hand side provide upper and lower bounds
(Bonferroni inequalities, see \cite{Fe}),
   i.e.\ for any   $n\leq [l/2]$:
$$
\sum_{k=0}^{2n} \sum_{{\AA_k=(i_1,\ldots, i_k) \atop
    \subset \{1,\ldots, l\} }
    \atop i_1<i_2<\cdots<i_k}
    {(-1)^k}\P\Big( \bigcap_{j=1}^k \bar O_{i_j}\Big)
\geq
 \P\Big(\bigcap_{ i=1,\ldots,l}O_i\Big)
   \geq \sum_{k=0}^{2n+1} \sum_{{\AA_k=(i_1,\ldots, i_k)
   \atop \subset \{1,\ldots, l\} }
    \atop i_1<i_2<\cdots<i_k}
    {(-1)^k}\P\Big( \bigcap_{j=1}^k \bar O_{i_j}\Big).
    \Eq(inex1)
$$
They imply that it will be enough to compute the limits as
$N\uparrow\infty$
of terms for any  fixed value of  $l$.
   Using \eqv(inex), we derive from the assumption of the theorem
   the following more general statement:
   Let $A_1, \ldots, A_l\in \R_+^p$ be any subsets
    of volumes $|A_1|,\ldots,
     |A_l|$ that can be represented as unions of disjoint
   cubes.  Then for any
   $m_1,\ldots, m_l$
$$
\sum_{\a(m_1),\a(m_2),\ldots,\a(m_l)}
     \P(\vec V_{i,M} \in A_j \ \forall i\in\a(m_r), \forall
     r=1,\ldots, l) \to \prod_{r=1}^l |A_r|^{m_r}.
\Eq(qq)
$$
    Let us first concentrate on the proof of this statement.
   We first show it in the case of one subset, $l=1$,
   which is a cube $A=\prod_{\b=1}^p [a^\b, b^\b)$.
   Let $m=1$.  We denote by  $\sum_{\AA\subset \{1,\ldots, p\}}$
   the sum over all $2^p$
   possible ordered  subsets of coordinates :
   $\AA$ denotes the subset of coordinates
   $\beta$  such that the inequalities
   $V_{i,M}^{\b}<a^{\b}$ are excluded
   leaving thus $V_{i,M}^{\b}<b^{\b}$.
   Then by \eqv(inex) applied to $\bigcap_{\b=1}^p
   \{V_{i,M}^{\b}\geq a^{\b}\}$
$$
\eqalign{\sum_{i=1}^{M} \P(\vec V_{i,M}\in A)=&
   \sum_{i=1}^{M}\P(a^{\b}\leq V_{i,M}^\b<b^{\b},
   \forall\b=1,\ldots,p)\cr
=&\sum_{i=1}^{M}\sum_{\AA \subset \{1,\ldots, p\}}(-1)^{|\AA|}
       \P(V_{i,M}^\b<a^{\b}\1_{\b\not\in \AA}+b^{\b}\1_{\b\in \AA},
   \forall \b=1,\ldots,p)\cr
=&\sum_{\AA \subset \{1,\ldots, p\}}(-1)^{|\AA|}
 \sum_{i=1}^{M}  \P(V_{i,M}^\b<a^{\b}\1_{\b\not\in \AA}+b^{\b}\1_{\b\in
 \AA},\forall \b=1,\ldots,p).
}
\Eq(tt)
$$
        The interior sum in \eqv(tt)
   $\sum_{i=1}^M \P(\cdot)$  converges to
  $\prod_{\b=1}^p( a^{\b}\1_{\b\not\in \AA}+b^{\b}\1_{\b\in
 \AA})$ by the assumption \eqv(maincond).
     Thus
$$\lim_{M\to \infty}
   \sum_{i=1}^{M} \P(\vec V_{i,M}\in A)=
   \sum_{\AA \subset \{1,\ldots, p\}}(-1)^{|\AA|}
   \prod_{\b=1}^p( a^{\b}\1_{\b\not\in \AA}+b^{\b}\1_{\b\in
 \AA})=\prod_{\b=1}^p (b^{\b}-a^{\b})=|A|.
\Eq(i)
$$
  Now let $m>1$.
  Denote by $\sum_{\AA_1,\AA_2,\ldots, \AA_m}$
  the sum over all $2^{mp}$ ordered sequences
  of all $2^p$  unordered subsets $\AA\subset\{1,\ldots, p\}$.
      Here $\AA_j$ is the subset of coordinates
  corresponding to the $j$th index
  in the row $\a(m)=(i_1,\ldots, i_m)$.
 Then by \eqv(inex)
 $$
\eqalign{
&\sum_{\a(m)} \P(\vec V_{i,M}\in A\ \forall i\in \a(m))
=\sum_{\a(m)}\P\left( a^{\b}\leq V_{i,M}^\b<b^{\b}\ \forall i\in
\a(m),\forall \b=1,\ldots,p\right)\cr
&=
\sum_{\a(m)}\sum_{\AA_1,\ldots, \AA_m}(-1)^{|\AA_1|+\cdots |\AA_m|}
       \P\left(V_{i,M}^\b<a^{\b}\1_{\b\not\in \AA_j}+b^{\b}\1_{\b\in \AA_j}
   \;\forall i=i_j\in \a(m),
\forall j=1,\ldots,m,
\forall \b\right)\cr
&=\sum_{\AA_1,\ldots, \AA_m}(-1)^{|\AA_1|+\cdots |\AA_m|}
 \sum_{\a(m)}\P(V_{i,M}^\b<a^{\b}\1_{\b\not\in \AA_j}+b^{\b}\1_{\b\in
 \AA_j}\, \forall i=i_j\in \a(m),
  \forall j=1,\ldots,m, \forall \b).
}
\Eq(tt1)
$$
By \eqv(maincond) applied to the interior sum of \eqv(tt1)
  $\sum_{\a(m)} \P(\cdot)$  we get:
$$\lim_{M\to \infty} \sum_{\a(m)} \P(\vec V_{i,M}\in A\ \forall i\in
\a(m))=\sum_{\AA_1,\ldots, \AA_m}(-1)^{|\AA_1|+\cdots |\AA_m|}
          \prod_{j=1}^m \prod_{\b=1}^p
(a^{\b}\1_{\b\not\in \AA_j}+b^{\b}\1_{\b\in
 \AA_j})=|A|^m.$$
    Assume now that $l>1$ and $A_r=\prod_{\b=1}^p [a^\b_r, b^\b_r)$,
    $r=1,\ldots,l$. Then
$$
\eqalign{
&\sum_{\a(m_1),\a(m_2),\ldots,\a(m_l)}
     \P\left(\vec V_{i,M} \in A_j \ \forall i\in\a(m_r), \forall
        r=1,\ldots, l\right)\cr
&= \sum_{\a(m_1),\a(m_2),\ldots,\a(m_l)}
    \sum_{\AA_1^{1},\ldots, \AA^{1}_{m_1},\atop
     ,\ldots, \AA_{1}^{l},\ldots,\AA^{l}_{m_l}}
      (-1)^{|\AA_1^1|+\cdots+ |\AA_{m_l}^l|}
    \P\bigl(V_{i,M}^\b<a^{\b}\1_{\b\not\in \AA_j^{r}}+b^{\b}\1_{\b\in
 \AA_j^r}\cr
&\qquad\qquad\qquad\qquad \qquad \forall i=i_j\in \a(m_r),
  \forall j=1,\ldots,m_r, \forall r=1,\ldots, l,
    \forall \b\bigr)\cr
&=\sum_{\AA^1_1,\ldots,\AA^1_{m_1}}
   (-1)^{|\AA^1_1|+\cdots+|\AA^1_{m_1}|}\cdots
   \sum_{\AA^l_1,\ldots,\AA^l_{m_l}}
   (-1)^{|\AA^l_1|+\cdots+|\AA^l_{m_l}|}
    \sum_{\a(m_1),\a(m_2),\ldots,\a(m_l)}
\cr
 &\quad \P\left(V_{i,M}^\b<a^{\b}\1_{\b\not\in \AA_j^{r}}+b^{\b}\1_{\b\in
 \AA_j^r}\, \forall i=i_j\in \a(m_r),
  \forall j=1,\ldots,m_r, \forall r=1,\ldots, l,
    \forall \b\right).
}
\Eq(gg)
$$

 Due to \eqv(maincond) applied once more to the interior sum
   $\sum_{\a(m_1),\ldots, \a(m_l)}\P(\cdot)$,
   \eqv(gg) converges to
$$
 \sum_{\AA^1_1,\ldots,\AA^1_{m_1}}
   (-1)^{|\AA^1_1|+\cdots+|\AA^1_{m_1}|}\cdots
   \sum_{\AA^l_1,\ldots,\AA^l_{m_l}}
   (-1)^{|\AA^l_1|+\cdots+|\AA^l_{m_l}|}
   \prod_{r=1}^{l}\prod_{j=1}^{m_r}
    \prod_{\b=1}^{p}( a^{\b}\1_{\b\not\in \AA_j^{r}}+b^{\b}\1_{\b\in
 \AA_j^r}) $$
 $$=\sum_{\AA^1_1,\ldots,\AA^1_{m_1}}
   (-1)^{|\AA^1_1|+\cdots+|\AA^1_{m_1}|}\cdots
   \sum_{\AA^{l-1}_1,\ldots,\AA^{l-1}_{m_{l-1}} }
   (-1)^{|\AA^{l-1}_1|+\cdots+|\AA^{l-1}_{m_{l-1}}|} $$
$$
\hphantom{zzzzzzzzzz} \prod_{r=1}^{l-1}\prod_{j=1}^{m_r}
    \prod_{\b=1}^{p}
( a^{\beta} \1_{ \beta\not\in \AA_{j}^{r}  }
+ b^{\b}\1_{ \b\in
 \AA_j^r } )|A_l|^{m_l} $$
$$=|A_1|^{m_1}|A_2|^{m_2}\cdots |A_l|^{m_l}.\hphantom{zzzzzzzzzzzzzzz
zzzzzzzzzzzzzzzzzz}
\Eq(ff)
$$

  Let finally  $A_1 =\bigcup_{k=1}^{s_1} A_{1,k},
  \ldots,A_l=\bigcup_{k=1}^{s_l} A_{l,k}$ be  unions of
  $s_1,\ldots, s_l$ disjoint
  cubes respectively.  Then we may write:
  $$
\eqalign{
& \sum_{\a(m_1),\a(m_2),\ldots,\a(m_l)}
     \P(\vec V_{i,M} \in A_j \ \forall i\in\a(m_r), \forall
        r=1,\ldots, l)\cr
&=\sum_{m_{1,1},\ldots, m_{1,s_1} \geq 0
        \atop
       m_{1,1}+\cdots +m_{1,s_1}=m_1}
   \cdots \sum_{m_{l,1},\ldots, m_{l,s_l} \geq 0
        \atop
       m_{l,1}+\cdots +m_{l,s_l}=m_l}
    \sum_{\a(m_{1,1}),\ldots, \a(m_{1, s_1}),\atop
         \ldots, \a(m_{l,1}),\ldots ,\a(m_{l,s_l})}\cr
&\quad
    \P\left( \vec V_{i,M} \in A_{r,k}    \forall i\in \a(m_{r,k})\;
    \forall r=1,\ldots,l,\forall k=1,\ldots, s_r\right)
}
\Eq(dd)
$$
  and apply to the interior sum
  $\sum_{\a(m_{1,1}),\ldots,  ,\a(m_{l,s_l}) }\P(\cdot)$
the statement \eqv(qq)
   about  cubes just proven by \eqv(ff). Then \eqv(dd)
  converges to
$$
\eqalign{
\sum_{m_{1,1},\ldots, m_{1,s_1} \geq 0
        \atop
       m_{1,1}+\cdots +m_{1,s_1}=m_1}
   \cdots \sum_{m_{l,1},\ldots, m_{l,s_l} \geq 0
        \atop
       m_{l,1}+\cdots +m_{l,s_l}=m_l}
    \prod_{r=1}^{l} \prod_{k=1}^{s_r}
          |A_{r,k}|^{m_{r,k}}
&= \prod_{r=1}^{l}\sum_{m_{r,1},\ldots, m_{r,s_r} \geq 0
        \atop
       m_{r,1}+\cdots +m_{r,s_r}=m_r}
    \prod_{k=1}^{s_r}  |A_{r,k}|^{m_{r,k}}
\cr&    =\prod_{r=1}^{l} |A_r|^{m_r}.
}
\Eq(AB.1005)
$$
  This finishes the proof of the statement \eqv(qq).

%       Finally, since any subset of $\R^p_+$ may be
%   approximated by unions of disjoint cubes , the statement
%  \eqv(qq) holds for any $A_1,\ldots, A_l$.

      Now we are ready to  turn to the proof of the theorem.
  The condition (i) has been already shown by \eqv(i).
  To verify (ii), let us construct a  cube
   $B=\prod_{\b=1}^p [0, \max_{l=1,\ldots, L} b_l^{\b})$
   of volume $|B|$,   then clearly $A\subset B$.
   For any $R>0$ we may write the following decomposition:
$$
\eqalign{
\P(\Pi_M(A)=0)&=\sum_{r=0}^{R}
   \frac{1}{r !} \sum_{\a(r)}
   \P\left( \vec V_{i,M}\in B\setminus A\, \forall i\in \a(r),
            \ \vec V_{i,M}\not \in B\, \forall
    i\not\in \a(r)\right)\cr
&
  + \P(\Pi_M(A)=0,\Pi_M(B)>R)\equiv I_1(R,M)+I_2(R,M).
 }
\Eq(AB.1006)
$$

 Applying the inclusion-exclusion \eqv(inex1)  principle to $M-r$  events
   $\{\vec V_i\not \in B \}$ for $i\not \in \a(r)$, we may  bound
 $I_1(R,M)$ for all $n\leq [(M-r)/2]$ by
 $$
\eqalign{
&\sum_{r=0}^{R}
   \frac{1}{r!}
\sum_{k=0}^{2n}\frac{(-1)^k}{k!}
  \sum_{\a(r),\a(k)}
   \P( \vec V_{i,M}\in B\setminus A \;\forall i\in \a(r), \
 \vec V_{i,M}\in B  \; \forall i\in \a(k))
\geq I_1(R,M)\cr
&\geq  \sum_{r=0}^{R}
   \frac{1}{r !}
\sum_{k=0}^{2n+1}\frac{(-1)^k}{k!}
  \sum_{\a(r)\a(k)}
   \P( \vec V_{i,M}\in B\setminus A \;\forall i\in \a(r),\
 \vec V_{i,M}\in B\; \forall i\in \a(k)).
}
\Eq(ss)
$$
  Then for any fixed $n\geq 1$,  the statement \eqv(qq)
  applied to the subsets $A/B$ and $B$ imply:
$$
\sum_{r=0}^{R}
   \frac{|B\setminus A|^r}{r!}
\sum_{k=0}^{2n}\frac{(-1)^k |B|^k}{k!}
   \geq \lim_{M\to \infty}
   I_1(R,M) \geq
  \sum_{r=0}^{R}
   \frac{|B\setminus A|^r}{r!}
\sum_{k=0}^{2n+1}\frac{(-1)^k |B|^k}{k!}.
\Eq(AB.1007)
$$
   Since $n$ can  be fixed arbitrarily large, it follows that
$$
\lim_{M\to \infty} I_1(R,M)=
  e^{-|B|}\sum_{r=0}^{R}\frac{|B\setminus A|^{r}}{r!}.
\Eq(hh)
$$
  The statement \eqv(qq) also gives
$$
\lim_{M\to \infty} I_2(R,M)
  \leq \lim_{M\to \infty} \P(\Pi_M^1(B)> R)=
   \lim_{M\to \infty} \frac{1}{R!}\sum_{\a(R)}
     \P(\vec V_{i,M}\in B \;\forall i\in \a(R))=
     \frac{|B|^{R}}{R!}.
\Eq(hhh)
$$
   By choosing $R$ large enough,
   the limit \eqv(hhh) can  be done as small
   as desired and
   the sum \eqv(hh) can  be done  as close to the exponent
   $e^{|B\setminus A|-|B|}$
   as wanted. Hence,
$\lim_{M\to \infty}\P(\Pi_M^1(A))=e^{-|A|}$.
   This concludes the proof of the theorem.
\endproof

\bigskip

\chap{3. Application to number partitioning}3
We will now prove Theorem \thv(1.1th). In fact, the proof will follow
 directly from
 Theorem \thv(mainth) and the following proposition:

\proposition{\TH(mainl)} {\it Let
$$
S(k,N)=  k^N  (2\pi N)^{\frac{1-k}{2}}
k^{\frac{k}{2}} (k!)^{-1}
\Eq(ck)
$$
  be borrowed from \eqv(sin).
     We denote by $\sum_{\s^1,\ldots,\s^l \in \S_N}(\cdot)$ the sum
  over all possible ordered sequences of {\sl  different} elements
   of $\S_N$.   Then for any $l=1,2,\ldots,$
  any constants $c_j^{\b}> 0$, $j=1,\ldots, l$,
   $\b=1,\ldots, k-1$ we have:

$$
\sum_{\s^1,\ldots, \s^l \in \S_N   }
  \P\Big( \forall \b=1,\ldots, k-1, \forall j=1,\ldots, l
      \frac{ |Y^{\b} (\s^j)| }{ \sqrt{ 2(N/k) \hbox{\rm var}\,X } }
     < \frac{ c_j^{\beta} } {S(k,N)^{\frac{1}{k-1}}  } \Big) $$
$$  \to  \prod\limits_{j=1,\ldots,l \atop \b=1,\ldots, k-1}
          (2(2\pi)^{-1/2}c_{j}^{\beta}).
\Eq(jk)
$$
 }

\noindent{\bf Informal arguments.} Before proceeding with
   the rigorous
  proof, let us give  intuitive arguments supporting  this lemma.

The random variables
   $\frac{ Y^{\b} (\s^j) }{ \sqrt{ 2(N/k) \hbox{\rm var}\,X } }$
are the sums of independent identically distributed random variables with
    the  expectations   $\E Y^{\b} (\s^j)  =0$ and  the covariance matrix
  $B_N(\s^1,\ldots,\s^l)$  with the elements
$$
 b^{\b,\g}_{i,s}= \frac{\hbox{\rm cov}\, (Y^\b(\s^i), Y^\g(\s^s))}{
2(N/k)
 \hbox{\rm var}\, X }
=  \frac{\sum_{n=1}^{N}
    (\1_{\{\s_n^i=\b\}}-\1_{\{\s_n^i=\b+1\}})
    (\1_{\{\s_n^s=\g \}}-\1_{\{\s_n^s=\g+1\}})}
{2(N/k)}.
\Eq(2a)
$$
 In particular:
$$
  b_{i,i,}^{\b,\b}=1,\ \
  b_{i,i}^{\b,\b+1}=-1/2,\ \
   b_{i,i}^{\b,\g} =0 \hbox{    for }\g\ne \b, \b+1,\ \forall \,i=1,\ldots,k-1.
\Eq(5a)
$$
 Moreover,  the property that
      $b^{\b,\g}_{i,j}=o(1)$ as $N\to \infty$ for all
   $i\ne j$, $\b$, $\g$,   holds  for
  a number $R(N,l)$ of sets $\s^1,\ldots, \s^l\in \Sigma_N^{\otimes l}$
   which is $R(N,l)=|\Sigma_N|^l(1+o(1))=
    S(k,N)^{l}(1+o(1))$ with $o(1)$ exponentially small
        as $N\to \infty$.
      For all such sets $\s^1,\ldots, \s^l$, by the Central Limit
      Theorem,
    the random variables
 $\frac{ Y^{\b} (\s^j) }{ \sqrt{ 2(N/k) \hbox{\rm var}\,X } }$
   should  behave asymptotically as centered  Gaussian random
 variables
  with covariances $b^{\b,\g}_{i,j}=1_{\{i=j, \b=\g\}}+
  (-1/2)1_{\{i=j, \g=\b+1\}}+o(1)$.
   The  determinant of this covariance matrix  is~$1+o(1)$.
        Hence, the probability $\P(\cdot)$ defined
           in \eqv(jk) that these Gaussians belong
  to the exponentially
       small segments\hfill \break $[-c_{j}^{\beta} S(k,N)^{-1/(k-1)},
 c_{j}^{\beta} S(k,N)^{-1/(k-1)}]$
   is of the order
    $\prod\limits_{j=1,\ldots,l \atop \b=1,\ldots, k-1}
           (2(2\pi)^{-1/2}c_{j}^{\beta} S(k,N)^{-1/(k-1)})$.
 Multiplying this probability by the number of terms  $R(N,l)$ we get
  the  result claimed in \eqv(jk).

    Let us turn to the remaining tiny part
  of $\Sigma_N^{\otimes l}$ where $\s^l,\ldots,\s^l$
  are such that $b^{\b,\g}_{i,j} \not \to 0$  for some $i\ne j$
  as $N\to \infty$.
      Here two possibilities should
  be considered differently. The first one is when
  the covariance matrix $B_N(\s^1,\ldots, \s^l)$ of
   $\frac{ Y^{\b} (\s^j) }{ \sqrt{ 2(N/k) \hbox{\rm var}\,X } }$
  is non-degenerate.
      Then invoking again the Central Limit Theorem,
   the probability $\P(\cdot)$  in this case is of the order
     $$(\hbox{\rm det} B _N(\s^1, \ldots, \s^l))^{-1/2}
             \prod\limits_{j=1,\ldots,l \atop \b=1,\ldots, k-1}
          (2(2\pi)^{-1/2}c_{j}^{\beta} S(k,N)^{-1/(k-1)}).$$
    But from the definition of $b_{i,j}^{\b, \g}$
        $(\hbox{\rm det} B_N(\s^1, \ldots, \s^l))^{-1/2}$
    may grow at most polynomially.
            Thus the probability $\P(\cdot)$
  is about $S(k, N)^{-l} $ up to a polynomial term  while the
        number of sets $\s^1, \ldots, \s^l$ in this part
   is exponentially smaller than $S(k, N)^{l}$.
   Hence, the contribution of all such $\s^l,\ldots,\s^l$
   in \eqv(jk) is exponentially small.

   The  case of $\s^1, \ldots, \s^l$
   with $B(\s^1,\ldots, \s^l)$ degenerate is more delicate.
        Although the number of such  $\s^1, \ldots, \s^l$
  is exponentially smaller than $S(k, N)^{l}$,
  the probability $\P(\cdot)$ is exponentially bigger than
    $S(k, N)^{-l}$ since the system
   of $l(k-1)$
     random variables $\{Y^{\b}(\s^i)\}_{\b=1,\ldots, k-1}^{
        i=1,\ldots,l}$
        is linearly dependent!
     First of all, it may happen that there exist  $1\leq
    i_1<i_2<\cdots <i_p \leq l$
     such that
    the basis of this system consists of $(k-1)p$ elements
      $\{Y^{\b}(\s^{i_j})\}_{\b=1,\ldots, k-1}^{j=1,\ldots,p}$.
     Then  the assumption that the elements $\s^1, \ldots, \s^l$
    of $\Sigma_N$  must be different, plays a crucial role:
     due to it the number of such sets  $\s^1, \ldots,\s^l$
    in this sum remains small enough
   compare to the probability $\P(\cdot)$, consequently
   their total contribution to \eqv(jk) vanishes.

    Finally, for some sets $\s^1,\ldots, \s^l$, there is no such $p<l$:
    for any basis, there exists a number $j\in \{1,\ldots, l\}$
   such that the random variables
     $Y^{\b}(\s^j)$ are included in the basis for some
   {\it non-empty\/} subset of coordinates
    $\b$ and are not included there for the complementary
   {\it  non-empty\/} subset of $\b$. This last part
    is clearly absent in the case $k=2$. It turns out that
   its analysis is quite tedious. We manage to complete
   it only in the case of the constrained problem by
   evaluating the number of such sets $\s^1, \ldots, \s^l$
   where each of spins' values
   $\{1,\ldots, k\}$  figures out exactly  $N/k$ times and by
    showing that the corresponding probabilities
   $\P(\cdot)$ are negligible compare to this number.
      The only drawback that remains in the study of the
   unconstrained problem is precisely the analysis of this part.

\bigskip

\noindent{\it Proof of Proposition  \thv(mainl).}
In the course of the proof we will rely on four lemmata that will be
stated here but proven  separately in  Section 4.
  Let
$$
f^{\s^1,\ldots, \s^l}_N(\{t_{\b,j}\})
   = \E \exp \Big(\frac{i}{ \sqrt{2(N/k) \hbox{\rm var}\, X}}
\sum_{j=1,\ldots,l,\atop
   \b=1,\ldots,k-1}t_{\b,j} Y^{\b}(\s^j)\Big)
\Eq(ab.100)
$$
   be the characteristic function of the random  vector
$(2(N/k)
 \hbox{\rm var}\, X)^{-1/2}
 \{Y^{\b}(\s^j)\}_{j=1,\ldots,l, \atop \b=1,\ldots,k-1}$.
   Here $\vec t =\{t_{\b,j}\}_{\b=1,\ldots,k-1, \atop j=1,\ldots,l}$
   is the vector with $(k-1)l$ coordinates.
 Then
$$
 \P\Big( \forall \b=1,\ldots, k-1, \forall j=1,\ldots, l
      \frac{ |Y^{\b} (\s^j)| }{ \sqrt{ 2(N/k) \hbox{\rm var}\,X } }
     < \frac{ c_j^{\beta} } {S(k,N)^{\frac{1}{k-1}}  } \Big)
\Eq(invf)
$$
$$
= \frac{1}{(2\pi)^{l(k-1)}}
   \lim_{D\to \infty} \int \limits_{[-D,D]^{l(k-1)}}\!\!\!\!\!\!\!\!
   f_N^{\s^1,\ldots, \s^l } (\{t_{\b,j}\})\!\!\!\!\!\!
\prod\limits_{j=1,\ldots, l,\atop
  \b=1,\ldots, k-1}
 \!\!\!\!\!\! \frac{e^{it_{\b,j} c^{\b}_j S(k,N)^{\frac{-1}{k-1} } }
    -e^{-it_{\b,j} c^{\b}_j S(k,N)^{\frac{-1}{k-1} } } }{it_{\b,j}}
  dt_{\b,j}.
$$
             It will be convenient to have in mind the following
         representation throughout the proof.
   Any configuration
   $\s$ gives rise  to  $k-1$ configurations
    $\s^{(1)},\ldots, \s^{(k-1)} \in \{-1,0,1\}^N$
 such that
 $$
\s_n^{(\b)}=\1_{\{\s_n=\b\}}-\1_{\{\s_n=\b+1\}},\ \ n=1,\ldots, N.
\Eq(qmq)
$$
    We now define the $ N \times (k-1)$ matrix $C(\s)$
    composed of columns
 $\s^{(1)},\ldots, \s^{(k-1)}$.
     Then it is composed of types of $k$ rows
   of length $k-1$:
      $O_0=(1,0,\ldots,0)$, $O_1=(-1,1,0\ldots,0)$,
 $O_2=(0,-1,1,0,\ldots,0),\ldots,$
   $O_{k-2}=(0,\ldots,0,-1,1)$,
   $O_{k-1}=(0,\ldots,0,-1)$.
 They correspond to  spin values
      $1,2,\ldots, k$  respectively:
  if $\s_n=\b$, then the $n$th row of $C(\s)$ is
     $O_{\b-1}$.\note{The case $k=2$ is particular, since here $C(\s)$
 is the vector with elements $\pm 1$; i.e.\ in this case this
 reparametrisation just corresponds to passing from values $\{1,2\}$ to
 $\{-1,+1\}$.}
 Each of these $k$  rows is repeated $N/k$ times in the
 construction of $C(\s)$.
    Then
$$Y^{\b}(\s)=\sum_{n=1}^N X_n \s_n^{(\b)}.$$
    Let
$C(\s^1,\ldots,\s^l)$ be the $N\times (k-1)l$
   matrix composed by the columns\hfill\break
$\s^{1, (1)}, \s^{1,(2)}$,$\ldots$,$ \s^{1,(k-1)}$,$\s^{2,(1)}$,
  $\ldots, \s^{l, (k-1)}$.  Then it is easy to see that
 the function\hfill\break   $f^{\s^1,\ldots, \s^l}_N(\{t_{\b,j}\})$
  is the product of $N$ functions
   $$
 f^{\s^1,\ldots, \s^l}_N(\{t_{\b,j}\})=\prod_{n=1}^N
    \E \exp \Big(\frac{i X_n }{ \sqrt{2(N/k) \hbox{\rm var}\, X}}
    \{  C(\s^1,\ldots, \s^l) \vec t \}_n\Big)$$
      $$ = \prod_{n=1}^N
     \frac{\exp \Big(\frac{i}{ \sqrt{2(N/k) \hbox{\rm var}\, X}}
    \{  C(\s^1,\ldots, \s^l) \vec t\}_n\Big)-1}{i
    (\sqrt{2(N/k) \hbox{\rm var}\, X})^{-1}
    \{ C(\s^1,\ldots, \s^l)\vec t \}_n },
\Eq(rtt)
$$
 where  $\{  C(\s^1,\ldots, \s^l) \vec t \}_n$  is the $n$th
 coordinate
 of the product of the vector $\vec t=\{t_{\b,j}\}_{\b=1,\ldots,k-1,\atop
  j=1,\ldots, l}$ with the  matrix $C(\s^1,\ldots, \s^l)$.

        We will split the sum  of \eqv(jk) into two terms
 $$\sum_{\s^1,\ldots,
 \s^l\in \Sigma_N}\P(\cdot)= \sum_{\s^1,\ldots, \s^l \in \Sigma_N \atop
    \hbox{\rm rank}\, C(\s^1, \ldots, \s^l)=(k-1)l}\P(\cdot)+
    \sum_{\s^1,\ldots, \s^l\in \Sigma_N \atop
    \hbox{\rm rank}\, C(\s^1, \ldots, \s^l)<(k-1)l}\P(\cdot)
\Eq(dva)
$$
   and show that the first term converges to the right-hand
  side of \eqv(jk) while the second term converges to zero.

       We start with the second term
   in \eqv(dva) that we split
  into two parts
$$
 \sum_{\s^1,\ldots, \s^l \in \Sigma_N  \atop
    \hbox{\rm rank}\, C(\s^1, \ldots, \s^l)<(k-1)l}\P(\cdot)
     =J_N^1+J_N^2.
\Eq(dva1)
$$
 In the first part $J_N^1$ the sum is taken
  over ordered sets $\s^1,\ldots, \s^l$ of different elements of
  $\Sigma_N$ with the following property:  the
   rank $r$  of $C(\s^1, \ldots, \s^l)$ is a multiple of
          $(k-1)$ and, moreover,
      there exist configurations
    $\s^{i_1}, \ldots, \s^{i_{r/(k-1)}}$
   such that all
   of $\s^{(1), i_1},\s^{(2), i_1},\ldots, \s^{(k-1),
  i_{r/(k-1)}}$ constitute the basis of the columns  of the matrix
      $C(\s^1, \ldots, \s^l)$, i.e.\ the rank of $C(\s^{i_1}, \ldots,
   \s^{i_{r/(k-1)}})$ equals $r$.
   Consequently, for any $j \in \{1,\ldots, l\}\setminus
   \{i_1, \ldots, i_{r/(k-1)}\}$ all of $\s^{(1), j},\ldots,
    \s^{(k-1),j}$ are linear combinations of the columns of the
  matrix $C(\s^{i_1}, \ldots,
   \s^{i_{r/(k-1)}})$.
 In the remaining part, $J_N^2$,
   the sum is taken over configurations $\s^1,\s^2,\ldots, \s^l$
    satisfying the complementary property:
    for any basis of the columns of
        $C(\s^{1}, \ldots, \s^{l})$
        there exist at least one configuration $\s^{i}$
       such that some of the configurations
    $\s^{(1), i}, \ldots, \s^{(k-1),i}$
   are included in this basis and some others are not\note{In the case
    $k=2$ the term $J_N^2$ can obviously not exist. This leads to
    considerable simplifications.}.

 The following  Lemma~\thv(pr2) shows that the sum $J_N^1$
  is taken over  sets  of different $\s^1, \ldots, \s^l$
   such that the matrix of the basis
      $C(\s^{i_1}, \ldots,
   \s^{i_{r/(k-1)}} )$  contains  at most
           $(k^{r/(k-1)}-1)$ different  rows.

    \lemma{\TH(pr2)} {\it Assume that the matrix $C(\s^1,\ldots,
  \s^l)$ contains all $k^l$ different  rows.
  Assume that a  configuration $\tilde \s$ is such that
       each  $\tilde \s^{(1)}, \ldots, \tilde \s^{(k-1)}$
   is a linear combination
   of the columns of the matrix $C(\s^1,\ldots, \s^l)$.
      Then the configuration $\tilde \s$ is obtained by a permutation
   of spin values  in one of
    the configurations $\s^{1},\ldots, \s^{l}$,
  i.e. $\tilde \s$
    coincides with one of $\s^1,\ldots, \s^l$ as an element
  of $\Sigma_N$.}

\remark In the case $k=2$, Lemma \thv(pr2) has been an important
  ingredient in the analysis of the Hopfield model. It possibly
  appeared first in a paper by Koch and Piasko \cite{KP}.

\medskip
  In fact, if in $J_N^1$
   the matrix $C(\s^{i_1}, \ldots, \s^{i_{r/(k-1)} })$ contained all
  $k^{r/(k-1)}$ different  rows,  then by Lemma~\thv(pr2)
   the remaining configurations  $\s^j$ with
     $j \in \{1,\ldots, l\}\setminus
   \{i_1, \ldots, i_{r/(k-1)}\}$
   would  be equal to one of
     $\s^{i_1}, \ldots,
   \s^{i_{r/(k-1)}}$ as elements of $\Sigma_N$, which is impossible
   since the sum in \eqv(dva1) is taken over {\it different\/} elements
   of $\Sigma_N$.  Thus
   there can be  at most $O((k^{r/(k-1)}-1)^N)$ possibilities
      to construct  $C(\s^{i_1}, \ldots, \s^{i_{r/(k-1)}})$ in the sum
        $J_N^1$.
   Furthermore, there is only a $N$-independent  number
      of possibilities to complete it by linear
   configurations of its columns up to
     $C(\s^1, \ldots, \s^l)$.
    To see this, assume that there are
       $\nu < k^{r/(k-1)}$
     different rows in the matrix
  $C(\s^{i_1}, \ldots, \s^{i_{r/(k-1)}})$
  and consider its restriction to these rows  which is
   the $\nu \times r$ matrix
 $\wt C(\s^{i_1}, \ldots, \s^{i_{r/(k-1)}})$.
  Then
     $\wt C(\s^{i_1}, \ldots, \s^{i_{r/(k-1)}})$
  has the same rank $r$ as  $C(\s^{i_1}, \ldots, \s^{i_{r/(k-1)}})$.
Now there are not more than $3^{(\nu (l(k-1)-r)) }$ ways to complete
the matrix $\wt C$ to a $\nu\times l(k-1)$ matrix with
   elements $1,-1,0$ such that all added columns
   of length $\nu$
   are linear combinations of those of $\wt C$.
 But each such choice determines uniquely the
  coefficients in these linear combinations, and hence the completion
   of the full $N\times r$ matrix
   $C(\s^{i_1}, \ldots, \s^{i_{r/(k-1)}})$ up to
  the  $N\times l(k-1)$
 matrix $C(\s^1,\ldots, \s^l)$ is already fully determined.
 Thus  the number of terms in the sum representing $J_N^1$ is smaller than
$$
 \sum_{\nu=r}^{k^{r/(k-1)}-1} \nu^N 3^{(\nu (l(k-1)-r)) }
=O (\big(k^{r/(k-1)}-1\big)^N).
\Eq(ab.1)
$$
     The next proposition gives  an a  priori estimate
  for each of these terms.

\lemma{\TH(pr1)} {\it There exists a constant $K(k,l)>0$
       such that for any different $\s^1,\ldots, \s^l \in \S_N$,
   any    $r=\hbox{\rm rank }C(\s^1,\ldots
\s^l)\leq (k-1)l$
 and all $N>1$
$$
 \P\Big( \forall \b=1,\ldots, k-1, \forall j=1,\ldots, l
      \frac{ |Y^{\b} (\s^j)| }{ \sqrt{ 2(N/k) \hbox{\rm var}\,X } }
     < \frac{ c_j^{\beta} } {S(k,N)^{\frac{1}{k-1}}  } \Big)
   \leq  K S(k,N)^{-r/(k-1)} N^{3r/2}.
\Eq(tra)
$$
}
      Hence,  by Lemma~\thv(pr1)
   each term in $J_N^1$ is  smaller than $K S(k,N)^{-r/(k-1)}N^{3r/2}$
   with the leading exponential term
   $k^{-Nr/(k-1)}$. It follows that
   $J_N^1=O\big( [(k^{r/(k-1)}-1)k^{-r/(k-1)}]^N\big) \to 0$ as $N\to \infty$.

  Let us now turn to $J_N^2$ in \eqv(dva1). The next proposition
  allows to evaluate the number of terms in this sum.

\lemma{\TH(pr3)} {\it  Let $D_N$ be any $N\times q$
             matrix of rank $r\leq q$.
          Assume that for any $N>1$
      it is composed only of $R$ different rows taken
     from a finite set $\DD$ of cardinality $R\geq k$.
          Let $Q_N (R,t)$ be the number
 of configurations $ \s$   such that
    the matrix $D_N$ completed by the columns
     $ \s^{(1)},\ldots, \s^{(k-1)}$
  has  rank $r+t$ where  $1\leq t\leq  k-2$.
      Then there exists a constant $K(R,t,k)>0$, depending only on
          $R,t,k$, such that
$$
 Q_N(R,t) \leq K(R,t,k) \frac{ (N(t+1)/k)!}{ ((N/k)!)^{t+1}}.
\Eq(pp.1)
$$
 }

  Now, to treat $J_N^2$, consider $\s^1, \ldots, \s^l$ such
     that  $(k-1)m+t_1+t_2+\cdots+t_s=r$
  columns of  $C(\s^1, \ldots, \s^l)$ form a basis for the span of all
     column vectors of this matrix.
     Then  there exist $\s^{i_1}, \ldots, \s^{i_m}$ such that
   all of $\s^{(v), i_p}$ are included in the basis  for all
     $v=1,\ldots, k-1$, $p=1,\ldots,m$,
      and there exist $\s^{j_1},\ldots, \s^{j_s}$
   such that among $\s^{(v), j_q}$
     $t_q \geq 1$ configurations are included in the basis
   and other $k-1-t_q \geq 1$ are not, $q=1,\ldots, s$.
          By Lemma~\eqv(pr3)
    the number of possibilities to construct
  such a matrix $C(\s^1, \ldots, \s^l)$
    is
$$
O\Big( k^{mN } \prod_{q=1}^s \frac{ (N(t_q+1)/k)!}{
  ((N/k)!)^{t_q+1}} \Big)\sim  k^{N m }\prod_{s=1}^q
      (t_q+1)^{N(t_q+1)/k}
$$
up to leading exponential order.
             The probability in~\eqv(dva) is already estimated
   in Lemma~\thv(pr1): it is
  $$
O(N^{3r/2} S(k, N)^{-r/(k-1)})\sim k^{-Nr/(k-1)}=
k^{-N(m(k-1)+t_1+t_2+\cdots +t_s)/(k-1)}.
$$
        Thus,  to conclude that $J_N^2 \to 0$
    exponentially fast,  it suffices to show  that for any $k=3,4, \ldots$
   and any $t=1,2,\ldots, k-2$ we have
   $ (t+1)^{(t+1)/k} k^{-t/(k-1)}<1$, which is reduced to
    the inequality
$$
\phi(k,t)=\frac{k-1}{t}\ln (t+1)-\frac{k}{t+1}\ln
  k<0.
$$
It is elementary to check that
   $\frac{\partial \phi(k,t)}{\partial k} <0$
  for all $k \geq t+1$ and $t\geq 1$.  Then, given $t$, it suffices
      to check this inequality for the smallest
  value of $k$  which is
      $k=t+2$,  that is that
   $$
\psi(k)=(k-1)^2\ln (k-1)-k(k-2)\ln k<0.
$$
     This is easy  as $\psi'(k)<0$ for all
   $k\geq 3$ and $\psi(3)<0$.
        Hence, $J_N^2\to 0$ as $N\to \infty$.
    Thus the proof of the convergence to zero of the second
  term of \eqv(dva) is complete.

        We now concentrate on the convergence of the
    first term of \eqv(dva).
 Let us fix any $\alpha\in (0, 1/2)$ and introduce
   a subset $\RR^{\a}_{l,N}\subset \Sigma_N^{\otimes l}$:
$$
\RR^{\a}_{l,N}= \Big\{\s^1, \ldots, \s^l \in \Sigma_N:
       \forall 1\leq i<r\leq l,
     1\leq \b,\gamma,\eta \leq k, \beta\ne \g$$
$$
 \Big|\sum_{n=1}^N (\1_{\{\s_n^i=\b\}}-\1_{\{\s_n^i=\gamma\}})
    \1_{\{\s_n^r=\eta\}}\Big|<N^{\a+1/2}
\Big\}.
\Eq(8.1)
$$
         This subset can be constructed as follows. Take $\s^1$
   where each of $k$ possible values of spins is present
  $N/k$ times. Divide each set  $\AA_{\b}\equiv \{i\in \{1,\dots,N\}:
   \s^1_i=\b \}$, $\b \in \{1,\dots,k\}$,  into
   $N/k+O(N^{\a+1/2})$ pieces $\AA_{\b,\g}$
       of length
    $N/k^2 +O(N^{\a+1/2})$.
           Then the spins of $\sigma^2$
    have the same value on the subsets
   of indices  which are composed
   by $k$ such pieces $\AA_{\b,\g}$  taken from different
   $\AA_\b$, $\b=1,\ldots, k$.
   Next,  divide $k^2$ subsets $\AA_{\b,\g}$ into $k$ pieces
    $\AA_{\b,\g,\delta}$.
   The spins of $\s^3$ have the same values on
   the subsets  composed by $k^2$  such pieces
    ${\cal A}_ {\b,\g,\delta}$  of
   length  $N/k^3 +O(N^{\a+1/2})$
  taken from different ${\cal A}_{\b,\g}$, etc.
    It is an easy combinatorial computation to check that
          with some constant $h>0$
 $$
|\Sigma_N^{\otimes l}\setminus \RR^{\a}_{l,N}|\leq
   k^{Nl}\exp(-h N^{2\a})
\Eq(cb)
$$
  from where by \eqv(sin)
$$
  |\RR^{\a}_{l,N}|= S(k,N)^l(1+o(1)), \ \ N\to \infty.
\Eq(rrre)
$$
    It is also not difficult to see that for any $\s^1, \ldots, \s^l \in
  \RR^\a_{l,N}$ the rank of $ C(\s^1, \ldots, \s^l)$ equals $(k-1)l$.
   Note  that the covariance matrix $B_N$  (see \eqv(2a))
can be expressed as
 $$
 B_N(\s^1,\ldots, \s^l)=\frac{C^T(\s^1,\ldots, \s^l) C(\s^1,\ldots, \s^l)}
     {2(N/k) \hbox{\rm var}\, X} .
\Eq(ttc)
$$
Thus by  definition of $\RR^{\a}_{N,l}$, its elements satisfy
$$
       b_{i,j}^{\b,\g}=O(N^{\a-1/2}) \ \ \ \forall\, \b,\g,
  i\ne j,
\Eq(df)
$$
uniformly for  $\forall\, \s^1,\ldots, \s^l \in \RR^{\a}_{N,l}$.
       Therefore, for any $\s^1,\ldots,\s^l \in \RR^\a_{l,n}$,
    $\hbox{\rm det}B_N(\s^1,\ldots,\s^l)=1+o(1)$
   and consequently the rank of $C(\s^1,\ldots,\s^l)$
   equals $(k-1)l$.

     By Lemma~\thv(pr1) and  the estimate \eqv(rrre)
 $$
\sum_{\s^1,\ldots, \s^l \not \in \RR_{l,N}^{\a} \atop
       \hbox{\rm rank} C(\s^1,\ldots, \s^l)=(k-1)l}\P(\cdot)  \leq
          k^{Nl}e^{-h N^{2\a}}K N^{3(k-1)l/2} S(k,N)^{-l} \to
          0.
\Eq(fg1)
$$
    To complete the study of the first
  term of \eqv(dva),  let us  show that
$$
 \sum_{\s^1,\ldots,\s^l \in \RR^\a_{l,N}}\P(\cdot)
   \to (2\pi)^{-(k-1)l/2} \prod\limits_{j=1,\ldots, l\atop
    \b=1,\ldots, k-1}(2c_j^\b)
\Eq(gh)
$$
  with $\P(\cdot)$ defined by \eqv(invf).
 Using the representation \eqv(invf),
will divide the normalized probability $\P(\cdot)$ of \eqv(invf)
     into five parts
$$
 S(k,N)^{l}
 \Big(\prod\limits_{i=1, \ldots, k-1, \atop j=1, \ldots, l}
    (2c_j^\b)^{-1}\Big)
\P(\cdot)=
   \sum_{i=1}^5 I_N^i(\s^1,\ldots,\s^l)
\Eq(55)
$$
where:
$$
I_N^1%(\s^1, \ldots, \s^l)
   \equiv   (2\pi)^{-l(k-1)}
  \int\limits_{\|\vec t\|<\epsilon N^{1/6}}
   e^{-\vec t B_N(\s^1, \ldots, \s^l) \vec t /2}
 \prod\limits_{\b=1, \ldots, k-1, \atop j=1, \ldots, l}
   d t_{\b,j},
\Eq(11)
$$
$$
I_N^2%(\s^1, \ldots, \s^l)
   \equiv   (2\pi)^{-l(k-1)}
  \int\limits_{\|\vec t\|<\epsilon N^{1/6}}
  \big(f_N^{\s^1,\ldots, \s^l}(\{t_{\b,j}\})
 - e^{-\vec t B_N(\s^1, \ldots, \s^l) \vec t /2}\big)
    \prod\limits_{\b=1, \ldots, k-1, \atop j=1, \ldots, l}
   d t_{\b,j},
\Eq(12)
$$
$$
I_N^3%(\s^1, \ldots, \s^l)
   \equiv  (2\pi)^{-l(k-1)}
  \int\limits_{\epsilon N^{1/6}<\|\vec t\|<\delta \sqrt{N}}
   f_N^{\s^1,\ldots, \s^l}(\{t_{\b,j}\})
 \prod\limits_{\b=1, \ldots, k-1, \atop j=1, \ldots, l}
   d t_{\b,j},
\Eq(13)
$$
$$
\eqalign{
 I_N^4%(\s^1, \ldots, \s^l)
\equiv &
(2\pi)^{-l(k-1)}\
 \int\limits_{\|\vec t\|\leq \delta \sqrt{N}}
    f_N^{\s^1,\ldots,\s^l}(\{t_{\b,j}\} )  \cr
&\times
  \Big[\prod\limits_{\b=1, \ldots, k-1 \atop j=1, \ldots, l}
       \frac{e^{i t_{\b,j} c_j^\b S(k,N)^{-1/(k-1)} }-
      e^{-i t_{\b,j} c_j^\b
 S(k,N)^{-1/(k-1) } }}{  2 i t_{j,\b} c_j^{\b} S(k,N)^{-1/(k-1)}
    }-1
 \Big]\prod\limits_{\b=1, \ldots, k-1, \atop j=1, \ldots, l}
    d t_{\b,j}
}
\Eq(14)
$$
and
 $$
\eqalign{
I_N^5%(\s^1, \ldots, \s^l)
\equiv&
   (2\pi)^{-l(k-1)} \lim_{D\to \infty}
   \int\limits_{[-D, D]^{l(k-1)}\cap \|\vec t\|>\delta \sqrt{N}}
   f_N^{\s^1,\ldots,\s^l} (\{t_{\b,j}\})
\cr
&\times \prod\limits_{\b=1, \ldots, k-1 \atop j=1, \ldots, l}
      \frac{e^{i t_{\b,j} c^\b_j S(k,N)^{-1/(k-1)} }-
      e^{-i t_{\b,j} c^\b_j S(k,N)^{-1/(k-1) } }}{ 2 i t_{\b,j} c_j^{\b}
    S(k,N)^{-1/(k-1)} }
    d t_{\b,j}.
}
\Eq(15)
 $$
for values $\d,\e >0$ to be chosen appropriately later.
We will show that there is a choice such
 that $I_N^{i}(\s^1,\ldots, \s^l) \to 0$ for
  $i=2,3,4,5$ and $I_N^1(\s^1, \ldots, \s^l) \to (2\pi)^{-(k-1)l/2}$,
   uniformly for
    $\s^1, \ldots, \s^l\in \RR_{l, N}^\a$ as $N\to \infty$.
    These facts combined  with \eqv(rrre) imply  the  assertion
   \eqv(gh) and complete the proof of the proposition.
    The following lemma  gives control over some of the terms
  appearing above.

\lemma{\TH(pr4)} {\it
  There exist  constants $C>0$, $\epsilon>0$, $\d>0$, and $\zeta>0$,
such that for all
     $\s^1, \ldots, \s^l \in \RR_{l,N}^\a$, the following estimates hold:
\item{(i)}
$$
\big |
 f_N^{\s^1,\ldots, \s^l}
(\{t_{\b,j}\})- e^{-\vec t B_N(\s^1, \ldots, \s^l) \vec t/2}
\big|\leq \frac{ C |\vec t\|^3}{\sqrt{N}}e^{-\vec t B_N(\s^1, \ldots, \s^l)
   \vec t/2}, \ \hbox{ for all }\|\vec t\|<\epsilon N^{1/6}
\Eq(p1)
$$
 \item{(ii)}
$$
\big |
  f_N^{\s^1,\ldots, \s^l}(\{t_{\b,j}\})\big| \leq
  e^{-\vec t B_N(\s^1, \ldots, \s^l) \vec t/2+ C |t|^3 N^{-1/2}}
  \ \hbox{ for all  }\|\vec t\|<\delta\sqrt{N},
\Eq(p2)
$$
and
\item{(iii)}     $$ \big |
  f_N^{\s^1,\ldots, \s^l}(\{t_{\b,j} \})\big| \leq e^{-\zeta
   \|\vec t\|^2}
    \ \hbox{ for all  }\|\vec t\|<\delta\sqrt{N}.
\Eq(p3)
$$
 }

We can now estimate the terms $I_N^i$.
   First, by a standard estimate on Gaussian integrals,
   $$
\eqalign{
I^1_N(\s^1,\ldots,\s^l)&=((2\pi)^{(k-1)l} \hbox{\rm det}
   B_N(\s^1,\ldots, \s^l))^{-1/2}+o(1)\cr
&= (2\pi)^{-(k-1)l/2}+o(1),\ \ N\to
   \infty,
}
\Eq(ab.100.1)
$$
   where $o(1)$ is uniform for $\s^1,\ldots, \s^l \in \RR_{l, N}^\a$
   by \eqv(df) and \eqv(5a).  Thus $I_N^1$ gives the desired main
   contribution.

        The second  part $I^2_N(\s^1,\ldots,\s^l)=
     O(N^{-1/2})$, uniformly
    for  $\s^1, \ldots, \s^l\in \RR_{l, N}^\a$
           by the estimates  \eqv(p1) and \eqv(df), \eqv(5a).
      The third part
     $I^3_N(\s^1,\ldots,\s^l)$  is exponentially small by \eqv(p3).
          To treat $I^4_N(\s^1,\ldots,\s^l)$,
     we note that for any $\epsilon>0$
   one can find $N_0$ such that for all $N\geq N_0$ and all
   $\vec t$ with
          $\|\vec t\|\leq \delta \sqrt{N}$
  the quantity in square brackets is smaller than $\e$
   in  absolute value,
      and apply again \eqv(p3).
  Finally, we estimate
$$
|I_N^5(\s^1, \ldots \s^l)|\leq
 (2\pi)^{-l(k-1)}
  \int\limits_{\|\vec t\|>\delta \sqrt{N}}|  f_N^{\s^1,\ldots,\s^l}
(\{t_{\b,j}\})| \prod_{j=1,\ldots,l,\atop
  \b=1,\ldots,k-1}
  dt_{\b,j}.
\Eq(16)
$$
  For any $\s^1,\ldots,\s^l \in
   \RR^\a_{l,N}$ the matrix
   $C(\s^1,\ldots, \s^l)$ contains all $k^l$ possible
  different  rows and
by \eqv(rtt)
     $ f_N^{\s^1,\ldots,\s^l}(\{t_{\b,j}\})$ is the product of $k^l$
         different characteristic
    functions, where each is taken
   to the power $N/k^l(1+o(1))$.
  Let us   fix from  a set of
   $k^l$   rows of $C(\s^1,\ldots,\s^l)$
            any $(k-1)l$ linearly independent
     and denote by $ \bar C$ the matrix composed by them.
  %  $\sqrt{\vec t \bar C \bar C^T \vec t}$  defines
  % a norm in $\R^{(k-1)l}$.
      There exists $\eta(\delta)>0$ such that
     $\sqrt {\vec t \bar C^T \bar C \vec t/
(2(1/k) \var X)}\geq \eta $ for all
    $\vec t$  with $\|\vec t\|>\delta$.
       Changing  variables
    $\vec s= \vec t\bar C^T  /\sqrt{2 (N/k) \var  X}$
 one gets the bound
$$
\eqalign{
|I_N^5(\s^1, \ldots, \s^5)|
   \leq& (2\pi)^{-l(k-1)}( 2(N/k) \var X)^{l(k-1)/2}
     (\det \bar C)^{-1}
\cr&
 \times  \int\limits_{\|\vec s\|>\eta }
     \prod\limits_{\b=1,\ldots, k-1, \atop  j=1,\ldots,l}
       \Big| \frac{e^{i s_{\b,j} }-1}{i s_{\b,j}}
  \Big|^{N k^{-l}(1+o(1))}
    ds_{\b,j}
\cr
&
 \leq C N^{l(k-1)/2}  (1-h(\eta))^{N k^{-l}(1+o(1))-2}
    \int\limits_{\|\vec s\|>\eta }
     \prod\limits_{\b =1,\ldots, k-1,\atop  j=1,\ldots,l}
   \Big| \frac{e^{i s_{\b,j} }-1}{i s_{\b,j}}
  \Big|^{2}
 ds_{\b,j},
}\Eq(18)
$$
 where $h(\eta)>0$ is chosen  such that
       $|(e^{is}-1)/s|<1-h(\eta)$ for all
    $s$ with $|s|>\eta/((k-1)l)$ and $C$ is a
   constant independent of the set
   $\s^1, \ldots, \s^l$ and $N$.
   Thus $I_N^5(\s^1,\ldots,\s^l)\to 0$,
   uniformly for $\s^1,\ldots, \s^l \in \RR^\a_{l,N}$,
   and exponentially fast as
   $N\to \infty$. This concludes  the proof
   of \eqv(gh) and of Proposition~\thv(mainl).
\endproof

\bigskip

\chap{4. Proofs of Lemmas  \thv(pr2), \thv(pr1), \thv(pr3),
       \thv(pr4).}4

\noindent{\it Proof of Lemma~\thv(pr2).}
     Let first $l=1$. Without loss of generality we may assume
      that the first $k$ rows of $C(\s^1)$ are different.
     Then for all $i=1,\ldots, k-1$,
   the following system of equations has a solution:
$$
\hphantom{a_1++} \lambda_1^{(i)}=\tilde  \s_1^{(i)}
$$
$$
  -\lambda_1^{(i)} +\lambda_2^{(i)} =\tilde \s_2^{(i)}
$$
$$
   -\lambda_2^{(i)} +\lambda_3^{(i)}=\tilde \s_3^{(i)}
$$
$$
\cdots
 $$
$$
-\lambda_{k-2}^{(i)}+\lambda_{k-1}^{(i)} =\tilde \s_{k-1}^{(i)}
$$
$$
\hphantom{ere\;}-\lambda_{k-1}^{(i)}=\tilde \s_k^{(i)}.
\Eq(sd)
$$
   Then necessarily  $\sum_{n=1}^{k} \tilde \s_n^{(i)}=0$ for all
     $i=1,2,\ldots,k-1$, since the sum of the left-hand sides
  of these equations equals $0$.
      But for at least one $j\in \{1,\ldots,k\}$
    and $i=1,\ldots, k-1$,
    $\s_{j}^{(i)}\ne 0$,
 for otherwise $\lambda_s^{(i)}=0$ for all $s=1,\ldots,k$
   $i=1,\ldots, k-1$
  and consequently  $C(\tilde \s)$ is composed
  only of zeros, which is impossible.
        Without loss of generality
  (by definition of $\Sigma_N$
     we may always permute spin values) we may
   assume that $\tilde \s_j^{(1)}\ne 0$.

We will use  the following crucial property of the configurations
  $\tilde \s^{(1)},\ldots,\tilde \s^{(k-1)}$:
 $$
\tilde \s_{n}^{(j)}=1 \Longrightarrow \tilde  \s_{n}^{(j+1)} =0,
   \tilde  \s_{n}^{(j+2)}=0,\ldots,
  \tilde  \s_{n}^{(k-1)}=0.
\Eq(hhn)
$$
$$
\tilde  \s_{n}^{(j)}=-1 \Longrightarrow \tilde \s_{n}^{(j+1)} =1,
   \tilde  \s_{n}^{(j+2)}=0,\ldots,
  \tilde  \s_{n}^{(k-1)}=0.
\Eq(hhhn)
$$
         It follows that, for a certain number $t_1\geq 1$
     of pairs of indices
    $n^1_1,n^2_1,\ldots, n^1_{t_1},n^2_{t_1}\in \{1,\ldots, k\}$, we
     must have that
  $\tilde \s_{n^1_u}^{(1)}=1$ and $\tilde
   \s_{n^2_u}^{(1)}=-1$, $u=1,\ldots,t_1$.
    We say that these $2t_1$  indices are ``occupied'' from the step
     $j=1$ on,
           since, by
      \eqv(hhn) and
\eqv(hhhn), we know all values
   $\tilde \s_{n^1_u}^{(j)}=0$ for all $j=2,3,\ldots, k-1$,
  $\tilde \s_{n^2_u}^{(2)}=1$, and
   $\tilde \s_{n^2_u}^{(j)}=0$ for all $j=3,\ldots, k-1$,
    $u=1,2,\ldots,t_1$. We say  that the  other
     $k-2t_1$ indices are ``free'' at step $j=1$.
          Then we must attribute to at least $t_1$ of
    $k-2t_1$ spins $\tilde \s_n^{(2)}$ with ``free''
      indices the value  $\tilde \s_{n}^{(2)}=-1$ in order to ensure that
           $\sum_{n=1}^{k}\tilde \s_{n}^{(2)}=0$.
     We could also attribute to a certain number $t_2\geq 0$ of pairs
  of the remaining $k-3t_1$ spins with ``free'' indices
   the values $\tilde \s_n^{(2)}=\pm 1$.
         Thus by \eqv(hhn), \eqv(hhhn) for $j=2$
    we know the values
  of $\tilde \s_n^{(j)}$ for $j=2,3,\ldots,k-1$
   for  at least $3t_1+2 t_2 $ indices $n$. We say that they
    are ``occupied'' from
   $j=2$ on.    Among them  $\tilde \s_n^{(3)}=1$  for
         the number of indices $t_1+t_2$ and
     $\tilde \s_n^{(3)}=0$ for the others $2t_1+t_2$.
   Then we should assign to the number $t_1+t_2$ of
   the remaining $k-3t_1-2t_2$ spins $\tilde \s_n^{(3)}$
      with ``free'' indices
   the value
    $\tilde \s_n^{(3)}=-1$
  to  make $\sum_{n=1}^k\s_n^{(3)}=0$.
   We could also attribute to a certain number  $t_3\geq 0$ of pairs
  of the remaining $k-4t_1-3t_2$ spins the values $\pm 1$.
     Hence, after the third step,
   $4t_1+3t_2+2t_3$ indices are ``occupied''  etc.
           Finally, after $(j-1)$th step,
     $jt_1+(j-1)t_2+\ldots +2t_{j-1}$ indices are ``occupied'',
     $\tilde \s_n^{(j)}=1$ for $t_1+\cdots +t_{j-1}$ among these
    indices, and
   at the $j$th step we  must put $\tilde \s_n^{(j)}=-1$
   for the same number $t_1+t_2+\ldots +t_{j-1}$ of ``free'' indices
    to ensure that  $\sum_{n=1}^k\tilde \s_n^{(j)}=0$.
       But, if
         $t_1>1$ or $t_1=1$ but  $t_{i}>0$
    for some  $2\leq i\leq k-1$, then, for some $j\leq k-1$, we have
   $$
k-jt_1-(j-1)t_2-\cdots -2t_{j-1}< t_1+t_2+\cdots+ t_{j-1}.
$$
   (In fact, for $j=k-1$,  if,  $t_1>1$, then
     obviously  $k-(k-1)t_1<t_1$, and if
   $t_1=1$ but $t_i>0$ we have $k-(k-1)-2<1$).
    This means that    at the  $j$th step
   there are  not enough ``free'' indices
        among the remaining $k-jt_1-(j-1)t_2-\ldots-2t_{j-1}$ ones
  such that we could assign $\tilde \s_{n}^{(j)}=-1$ to ensure
          $\sum_{n=1}^k \tilde \s_{n}^{(j)}=0$.
  Hence, the only possibility is $t_1=1$ and $t_2=t_3=\cdots= t_{k-1}=0$.

 So,  at the first step
      $2$ indices get ``occupied'' and at each step
   one more index is ``occupied''. Thus there exists a sequence
  of $k$ different indices $n_1, n_2,\ldots, n_k\in \{1,\ldots, k\}$
    such that
  $\s_{n_i}^{(i)}=1$, $\s_{n_{i+1}}^{(i)}=-1$,
     $\s_{n}^{(i)}=0$ for $n\ne n_i,n_{i+1}$, $i=1,\ldots, k-1$.
    Solving the system \eqv(sd), we see that
    $\lambda_{n_i}^{(i)}=\lambda_{n_i+1}^{(i)}=\cdots
    =\lambda_{n_{i+1}-1}=1$, $\lambda_{n}^{(i)}=0$ for
    $n\ne n_i,\ldots, n_{i+1}-1$. Hence,
      the configuration $\tilde \s$
  is a permutation of the configuration $\s^{1}$
     such that $\tilde \s_n=i$,  iff  $\s^1_{n_i}=i$,
       $i=1,\ldots,k$.

        Let us now turn to the case $l>1$. We use induction.
     Consider $k^{l-1}$ possible columns.
 We denote linear combinations of them by
  $\Lambda_\a^{(i)}$, $\a=1,\ldots, k^{l-1}$.
  Then for any $i=1,\ldots, k-1$, the following system should have a solution
$$
\eqalign{ \Lambda_\a^{(i)}+ \lambda_1^{(i)}&= \tilde
     \s_{1,\a}^{(i)}\cr
 \Lambda_\a^{(i)}
   -\lambda_1^{(i)} +\lambda_2^{(i)} &=\tilde \s_{2,\a}^{(i)}\cr
 \Lambda_\a^{(i)}  -\lambda_2^{(i)} +\lambda_3^{(i)}& =
\tilde \s_{3,\a}^{(i)}\cr
\cdots & =\cdots \cr
\Lambda_\a^{(i)}
  -\lambda_{k-2}^{(i)}+\lambda_{k-1}^{(i)} &=\tilde \s_{k-1,\a}^{(i)} \cr
\Lambda_\a^{(i)} -\lambda_{k-1}^{(i)}=&\tilde \s_{k,\a}^{(i)}.
}
\Eq(sd1)
$$
      It follows that
  $$
\eqalign{
2\lambda_1^{(i)}-\lambda_2^{(i)}=&\tilde \s_{1,\a}^{(i)}-
  \tilde \s_{2,\a}^{(i)}
\cr
 - \lambda_1^{(i)}+2\lambda_2^{(i)}-\lambda_3^{(i)}=&
     \tilde \s_{2,\a}^{(i)}-\tilde \s_{3,\a}^{(i)},
\cr
 \cdots = &\cdots \cr
 -\lambda_{k-2}^{(i)}+2\lambda_{k-1}^{(i)}
    =&\tilde \s_{k-1,\a}^{(i)}-\tilde \s_{k,\a}^{(i)}.
}
 \Eq(fg)
$$
 Given  $\tilde \s_{1,\a}^{(i)}-\tilde \s_{2,\a}^{(i)},\ldots,
    \tilde \s_{k-1,\a}^{(i)}-\tilde \s_{k,\a}^{(i)}$,
   this system \eqv(fg)
   of $k-1$ equations
       has a unique
    solution, which does not depend on $\a =1, \ldots ,k^{l-1}$.
   Then  $\tilde \s_{1,\a}^{(i)}-\tilde \s_{2,\a}^{(i)},\ldots,
        \tilde \s_{k-1,\a}^{(i)}-\tilde \s_{k,\a}^{(i)}$
     should not depend on
          $\a$ neither.
  We denote by $\delta_{j}^{(i)}=\tilde \s_{j,\a}^{(i)}-\tilde
   \s_{j+1,\a}^{(i)}$.

   Let us  consider two cases. In the first case we assume that,
for some $i=1,\ldots,k-1$ and
    for some $j=1,\ldots, k-1$,
   $\delta_{j}^{(i)}\ne 0$.
   Then it may take values $\pm 1, \pm 2$.
   Knowing each of these values, we can  reconstruct in a unique way
        $\tilde \s_{j,\a}^{(i)}=\tilde \s_{j}^{(i)}$ and
     $\tilde \s_{j+1, \a}^{(i)}=\tilde \s_{j+1}^{(i)}$, which do not
   depend on $\a$.
   (If   $\delta_{j}^{(i)}=1$, then
     $\tilde \s_{j}^{(i)}=1$ and $\tilde \s_{j+1}^{(i)}=0$,
     if  $\tilde \delta_{j}^{(i)}=-1$, then
     $\tilde \s_{j}^{(i)}=0$ and $\tilde \s_{j+1}^{(i)}=-1$ etc.).
   Then we can reconstruct the values $\tilde \s_{t,\a}^{(i)}
     =\tilde \s_{j}^{(i)}+\sum_{m=t}^{j-1} \delta_{m}^{(i)}$ for
      $t=1,\ldots,j-1$,
    $\tilde \s_{t,\a}^{(i)}
     =\tilde \s_{j}^{(i)}-\sum_{m=j}^{t-1} \delta_{j}^{(i)}$
       for $t=j+1,\ldots,k$,  which consequently
      do not depend on $\a$.
  Since the sum of all $k^l$ left-hand sides of
   equations \eqv(sd1) equals zero, it follows that $\sum_{\a}
    \sum_{j=1}^k \tilde \s_{j,\a}^{(i)}=0$.
   But, since  $\tilde \s_{j,\a}^{(i)}=\tilde \s_{j}^{(i)}$,
         it follows that  $\sum_{j=1}^k \tilde \s_{j}^{(i)}=0$.
    Thus, $\Lambda_{\a}^{(i)}
              = \frac{1}{k} \sum_{j=1}^k \s_{j,\a}^{(i)}=
          \frac{1}{k} \sum_{j=1}^k \tilde \s_{j}^{(i)}=0$ for all $\a$.

    The sequence $\tilde \s_{1}^{(i)},\ldots, \tilde \s_{k}^{(i)}$ being
         not constant and  $\sum_{j=1}^k \tilde \s_{j}^{(i)}=0$,
    it follows that for some $j_1,j_2$,
       $\tilde \s_{j_1}^{(i)}=1$ and
   $\tilde \s_{j_2}^{(i)}=-1$.  Using
   \eqv(hhn) and \eqv(hhhn), we see that
     $\tilde \s_{j_1}^{(i+1)}=0$ and  $\tilde \s_{j_2}^{(i+1)}=1$.
     Therefore, for some $j=1,\ldots,k-1$ $\delta_j^{(i+1)}\ne 0$, so
         that  we may apply the previous reasoning to
  the configuration $\tilde \s^{(i+1)}$. We  get that the values
    $\tilde \s_{j,\a}^{(i+1)}$  do not depend on $\a$ and that
  $\Lambda_{\a}^{(i+1)}=0$, for all $\a$.
    Applying the  analogues of \eqv(hhn) and \eqv(hhhn)  backwards, namely
 $$
\tilde \s_{n}^{(j)}=-1 \Longrightarrow \tilde \s_{n}^{(j-1)} =0,
    \tilde \s_{n}^{(j-2)}=0,\ldots,
   \tilde \s_{n}^{(1)}=0,
\Eq(hh1)
$$
$$
\tilde \s_{n}^{(j)}=1 \Longrightarrow \tilde
     \s_{n}^{(j-1)} =-1, \tilde \s_{n}^{(j-2)}=0,\ldots,
   \tilde \s_{n}^{(1)}=0,
\Eq(hhh2)
$$
  we find that
   $\tilde \s_{j_1}^{(i-1)}=-1$ and $\tilde \s_{j_2}^{(i-1)}=0$.
  Thus, for some $j=1,\ldots,k-1$, $\delta_j^{(i-1)}\ne 0$
  and so we may apply the previous reasoning to
   the configuration $\tilde \s^{(i-1)}$. Hence,
   $\tilde \s_{j,\a}^{(i-1)}$ does not depend on $\a$ and
  $\Lambda_{\a}^{(i-1)}=0$ for all $\a$.
   Continuing this reasoning
subsequently  for $\tilde \s^{(i+2)},\ldots,\tilde \s^{( k)}$
   and backwards for $\tilde \s^{(i-2)},\ldots, \tilde \s^{(1)}$,
   we derive that
  none of the values  $\tilde \s_{j,\a}^{(s)}$ depends on $\a$
  and that $\Lambda_{\a}^{(s)}=0$ for all $\a$ and
      all $s=1,\ldots,k$. But  the  system
   $\Lambda_{\a}^{(s)}=0$ for all $s=1,\ldots, k-1$ and
   $\a=1,\ldots, k^{l-1}$  has only the trivial solution.
  Hence the system \eqv(sd1) becomes the system \eqv(sd).
      Invoking the reasoning for $l=1$, we derive that
  $\tilde \s$ is a permutation of the last configuration~$ \s^l$.

    Let us now turn to the second case, that is   assume that for all $i,j$
  $\delta_j^{(i)}=0$. Then the unique solution of  \eqv(fg)
   is $\lambda_1^{(i)}=\cdots=\lambda_{k-1}^{(i)}=0$.
      Then $\tilde \s_{1,\a}^{(i)}=\tilde \s_{2,\a}^{(i)}=
   \cdots =\tilde \s_{k,\a}^{(i)}=
  \tilde \s_{\a}^{(i)}$ for all
   $\a=1,\ldots, k^{l-1}$ and all $i=1,\ldots,k-1$.
  The system \eqv(sd1) is reduced to a smaller system
$\Lambda_\a^{(i)}=\tilde \s_{\a}^{(i)}$ corresponding
  to the matrix $C(\s^1,\ldots, \s^{l-1})$
with all $k^{l-1}$ different columns.
 The statement of
   the lemma holds for it by induction.
      Thus in this case $\tilde \s$ is a permutation
  of one of $\s^1,\ldots , \s^{l-1}$.
\endproof

\noindent{\it Proof of Lemma~\thv(pr1).}
   Let us remove from the matrix $C(\s^1,\ldots, \s^l)$
       linearly dependent columns and leave only $r$
   columns of the basis.  They correspond to
   a certain subset of $r$ configurations $\s^{j, (\b)}$
   $j,\b \in  A_r \subset \{1,\ldots, l\} \times \{1,\ldots,
      k-1\}$, $|A_r|=r$.
    We denote  by $\bar C^r(\s^1,\ldots, \s^l)$
the $N\times r$ matrix composed by them.
      Then the probability in the right-hand side of \eqv(tra) is not
  greater than the probability of the same
  events for   $j, \b \in  A_r$ only.
    Let
    $\bar f_N^{\s^1,\ldots, \s^l}(\{t_{\b,j}\})$,
   $j,\b\in  A_r$, be the characteristic function  of the
  vector  $(2(N/k) \hbox{\rm var}\, X)^{-1/2}
   \{Y^{\b}(\s^j)\}_{j, \b \in  A_r}$. Then
$$
 \P\Big( \forall \b=1,\ldots, k-1, \forall j=1,\ldots, l
      \frac{ |Y^{\b} (\s^j)| }{ \sqrt{ 2(N/k) \hbox{\rm var}\,X } }
     < \frac{ c_j^{\beta} } {S(k,N)^{\frac{1}{k-1}}  } \Big)$$
$$ \leq  \frac{1}{(2\pi)^{r}}
   \lim_{D\to \infty} \int \limits_{[-D,D]^{r}}
   \bar f_N^{\s^1,\ldots, \s^l } (\{t_{\b,j}\})
\prod\limits_{j,\b \in {A}_r}
  \frac{e^{it_{\b,j} c^{\b}_j S(k,N)^{\frac{-1}{k-1} } }
    -e^{-it_{\b,j} c^{\b}_j S(k,N)^{\frac{-1}{k-1} } } }{it_{\b,j}
     }
  dt_{\b,j}.
\Eq(tt1.2)
$$
 To bound the integrand in \eqv(tt1.2) we use that
$$
 \Big| \frac{e^{it_{\b,j} c^{\b}_j S(k,N)^{\frac{-1}{k-1} } }
    -e^{-it_{\b,j} c^{\b}_j S(k,N)^{\frac{-1}{k-1} } }
 }{it_{\b,j}}\Big|
\leq  \min \Big( 2 c^{\b}_j S(k,N)^{\frac{-1}{k-1}},
    2(t_{\b,j})^{-1} \Big).
\Eq(AB.101)
$$
 Next, let us choose in the matrix $\bar C^{r}(\s^1,\ldots,\s^l )$
       any $r$ linearly independent  rows  and
    construct of them a $r\times r$
        matrix $\bar C^{r\times r}$. Then
$$
\eqalign{
|\bar f^{\s^1,\ldots, \s^l}_N(\{t_{\b,j}\})|=&\prod_{n=1}^N
    \Big|\E \exp \Big(\frac{i X_n }{ \sqrt{2(N/k) \hbox{\rm var}\, X}}
    \{  \bar C^r(\s^1,\ldots, \s^l) \vec t \}_n\Big)\Big|\cr
    \leq& \prod_{s=1}^r
    \Big|\E \exp \Big(\frac{i X_s }{ \sqrt{2(N/k) \hbox{\rm var}\, X}}
    \{  \bar C^{r\times r} \vec t \}_s\Big)\Big|\cr
\leq&
    \prod_{s=1}^r \min \Big(1,
    2\sqrt{2( N/k) \hbox{\rm var}\, X }
    (\{ \bar C^{r\times r}\vec t\}_s)^{-1}\Big),
}
\Eq(AB.102)
$$
  where $\vec t=\{t_{\b,j} \}_{j,\b \in A_r}$.
    Hence, the absolute value of the
      integral \eqv(tt1.2) is bounded by
 the sum of two terms
  $$
  \frac{ S(N,k)^{\frac{-r}{k-1}}
\prod_{\b,j\in A_r} (2c_{\b}^j)}{(2\pi)^{r}}
  \int\limits_{\|\vec t\|<S(k,N)^{\frac{1}{k-1}}}
   \prod\limits_{s=1}^r \min
    \Big(1,  2\sqrt{2( N/k) \hbox{\rm var}\, X }
    (\{ \bar C^{r\times r} \vec t \}_s)^{-1}\Big) dt_{\b,j}
$$
$$
{}+\frac{1}{(2\pi)r}
    \int\limits_{\|\vec t\|>S(k,N)^{\frac{1}{k-1}}}
         \prod_{\b,j \in A_r}( 2 (t_{\b,j})^{-1})
       \prod\limits_{s=1}^r \min
    \Big(1, 2 \sqrt{2( N/k) \hbox{\rm var}\, X }
    (\{ \bar C^{r\times r} \vec t \}_s)^{-1}\Big) dt_{\b,j}.
\Eq(yy)
$$
  The change of variables $\vec \eta=
    \bar C^{r\times r} \vec t $ in the first term
  shows  that the integral over
    ${\|\vec t\|<S(k,N)^{\frac{1}{k-1}}}$
  is at most
$O( N^{r/2}
  (\ln S(k,N)^{\frac{1}{k-1}})^r)$, where
   $\ln S(k,N)^{\frac{1}{k-1}}
  = O(N)$ as $N \to \infty$.
        Thus the first term of \eqv(yy) is bounded by
   $K_1(\bar C^{r\times r},k,l)N^{3r/2} S(k,N)^{\frac{-r}{k-1}}$
  with some constant $K_1(\bar C^{r \times r},k,l)>0$ independent of $N$.
  Using the change of variables
    $\vec \eta = S(k,N)^{\frac{-1}{k-1}}\vec t$ in the second term
  of \eqv(yy), one finds  that
    the integral over ${\|\vec t\|>S(k,N)^{\frac{1}{k-1}}}$
  is at most $O(S(k,N)^{\frac{-r}{k-1}})$.
  Thus \eqv(yy) is not greater than $K_2(\bar C^{r\times r}, k,l) N^{3r/2}
   S(k,N)^{\frac{-r}{k-1}}$ with some positive constant
      $K_2(\bar C^{r\times r}, k,l)$ independent of $N$.

       To conclude the proof,
  let us recall the fact that
  there is a finite, i.e.\ $N$-independent,  number
  of possibilities to construct the matrix $\bar C^{r\times r}$
  starting from $C(\s^1, \ldots, \s^l)$
    since each of its elements may take
  only three values $\pm 1, 0$.
  Thus there exists less than  $3^{r^2}$
    different constants $K_2(\bar C^{r \times r}, k, l)$
  corresponding to different matrices $\bar C^{r \times r}$.
  It remains to take the maximal one over them
  to get \eqv(tra).
\endproof

\noindent{\it Proof of Lemma~\thv(pr3).}
  Throughout the proof we denote by $D_N\cup C(\s)$ the matrix $D_N$
  completed by the rows $\s^{(1)},\ldots, \s^{(k-1)}$.

      Let us denote by $c^1,\ldots, c^q$ the system
   of columns of the matrix $D_N$.
     Then we can find the indices $i_1<i_2<\ldots<i_{k-t-1}\leq k-1$
  such that $\s^{(i_s)}$ is a linear combination of
  $c^1,\ldots, c^q, \s^{(1)}, \s^{(i_2)},\ldots, \s^{(i_s-1)}$
  for all $s=1,\ldots, k-t-1$.
 Then there exist linear coefficients
       $a_1(s),\ldots, a_{i_s-1}(s)$ such that
$$
  a_1(s) c^1+\cdots + a_q(s)c^q + a_{q+1}(s)\s^{(1)}+
   \cdots+ a_{i_s-1}(s)\s^{(i_s-1)}=\s^{(i_s)}, \ s=1,\ldots, k-t-1.
\Eq(abn)
$$
 (If $r<q$ these coefficients may be not unique, but this is
  not relevant for the proof.)  Since $t\geq 1$,
    without loss of generality (otherwise just make a permutation of
    spin values $\{1,\ldots, k\}$ in $\s$) we may assume that $i_1>1$.

  Initially
   each of $k-t-1$
    systems \eqv(abn) consists of  $N$ linear equations.
   But the number of different rows of $D_N$ being a fixed number  $R$,
   each  of these $k-t-1$  systems \eqv(abn) has only a finite number
   of {\it different} equations.
Thus, \eqv(abn) are equivalent to  $k-t-1$ {\it finite} (i.e.\ $N$-independent)
   systems of different equations  of the form:
$$
  a_1(s) d_1+\cdots + a_q(s)d_q = a_{q+1}(s)\delta_1+
   \cdots+ a_{i_s-1}(s)\delta_{i_s-1}+\delta_{i_s},
\Eq(abn1)
$$
       where $d=(d_1,\ldots, d_q)$ is one of the $R$ distinct   rows
       of the  matrix $D_N$
  and $\delta_j=0,1,-1$.

Note that there exist at most $R \times 3^s$
  of such  equations \eqv(abn1) for any $s=1,\ldots, k-t-1$.
Consequently, for the given matrix $D_N$,
 there exists a {\it finite\/} (i.e. $N$-independent)
 number  of such sets of $k-t-1$ finite systems of
 distinct equations \eqv(abn1).
   We will denote by  $\AA$ the set of such sets of $k-t-1$
 finite systems of distinct equations \eqv(abn1)  which
do arise from some choice of a  spin configuration $\s$  with
   $ \hbox{\rm rank }[D_N\cup C(\s)] = r+t$,
after the
 reduction of \eqv(abn) (i.e.\ after eliminating the same equations
   among all $N$ in each of $k-t-1$ systems \eqv(abn)).
For $\s\in\S_N$, we denote by $\a(\s)\in \AA$ the set of $k-t-1$ finite systems
   of distinct equations \eqv(abn1)
  obtained from \eqv(abn) in this way.

          We will prove that for any
  given element $\a \in \AA$ we have the estimate:
  $$
\#\{\s: \hbox{\rm rank} [D_N \cup C(\s)]=r+t,\ \  \a(\s)=\a\}\leq
    C \frac{(N(t+1)/k)!}{((N/k)!)^{t+1}}
\Eq(ik.1)
$$
where $C$ is a constant that depends only on $R,t,k$.
  Since the cardinality of  $\AA$ is finite and depends only on $R,t$,
  and $k$, this will prove the lemma.

       Consider some  $\a_0 \in \AA$. Since by definition of $\AA$
  there exists $\s_0$ with the property $\hbox{\rm rank} [D_N \cup
  C(\s_0)]=r+t$  and $\a(\s_0)=\a_0$, then
  there  exists a solution of all these $k-t-1$ systems of equations
  $\a_0$.  Let $a_i(s)$  be any such solution.
   For any row $d=(d_1,\ldots, d_q)$ of $D_N$, set
   $$
\Lambda(s,d)=a_1(s)d_1+a_2(s)d_2+\cdots a_q(s)d_q.
\Eq(ik.2)
$$
          Then to any row $d$ of $D_N$ there corresponds
  the vector of linear combinations
   $\Lambda(d)=(\Lambda(1,d),\Lambda(2,d),\ldots,
         \Lambda(k-1-t,d))$.
   Next,  let us divide the set
   $\DD$ of the  $R$ different rows of the matrix $D_N$ into
   $m$ disjoint non-empty subsets $\DD_1,\DD_2,\ldots, \DD_m$
such that two rows $d,\tilde d$ are in the same subset, if and only if
$\L(d)=\L(\tilde d)$.

\lemma{\TH(AB.1)}{\it The partition $\DD_i$ defined above satisfies
  the following properties:
\item{(i)}    $m\geq k-t$
\item{(ii)} For any pair $d \in \DD_i$, $\tilde d \in \DD_j$, with $i\ne j$,
            and for any $\s$, such that $\hbox{\rm  rank}[D_N \cup
        C(\s)] =r+t$  and
           $\a(\s)=\a_0$, the rows
       $d$ and $\tilde d$ can  not be continued by
       the same row $O$ of the matrix $C(\s)$ in $D_N \cup C(\s)$.
}

\noindent{\it Proof.}    Let us first show that $\DD$ can be divided
 into three non-empty subsets $\DD_0$, $\DD_1$,  $\DD_2$,
  such that $\Lambda(1,d) \ne -1,0$ for all $d\in \DD_0$,
     $\Lambda(1,d)=-1$ for $d\in \DD_1$,
         $\Lambda(1,d)=0$ for $d \in \DD_2$.
    First of all, since $\a_0 \in \AA$, then there exists
     at least one $\s_0$ such that
     $\hbox{\rm rank} [D_N \cup
  C(\s_0)]=r+t$  and $\a(\s_0)=\a_0$.
    Let  $d^0,\ldots, d^{k-1}$ denote $k$ rows (not necessarily different)
    of $D_N$ that are continued
    by the rows $O_0, \ldots, O_{k-1}$   of the matrix
       $C(\s_0)$ (recall the definition given in
  the paragraph following \eqv(qmq)) respectively in $D_N \cup C(\s_0)$.
Now consider a row  $d^{i_1}$  that is continued
    by the row $O_{i_1}$.
The corresponding equation \eqv(abn) with $s=1$ then reads
   $$
\Lambda(1,d^{i_1})=-1.
$$
This shows that the set $\DD_1\neq \emptyset$.
Similarly,  for a row $d^{j}$  continued
    by the row $O_{j}$  with $j>i_1$, the corresponding equation yields
      $$
\Lambda(1,d^j)=0.
$$
Thus $\DD_2\neq\emptyset$.
       Finally, consider the rows  of the matrix $D_N$
   continued by $O_0,\ldots, O_{i_1-1}$.
   The corresponding $i_1$ equations  \eqv(abn) with $s=1$ then read :
     $$
\eqalign{
\Lambda(d^0,1)&=-a_{q+1}\cr
\Lambda(d^1,1)&=a_{q+1}-a_{q+2}\cr
\Lambda(d^2,1)&=a_{q+2}-a_{q+3}\cr
\cdots&=\cdots\cr
\Lambda(d^{i_1-1},1)&=  a_{q+i_1-1}+1.
}
\Eq(abn.10)
$$
   The sum of the right-hand sides of these equations
     equals $1$. Thus the left-hand side of at least one
   equation must be  positive.  Hence,  there exists
   $d^j$ with $j\in \{1, \ldots, i_1-1\}$ such that
     $$
\Lambda(d^j,1)\ne -1,0.
$$
Thus also $\DD_0\neq\emptyset$, and so all three sets defined above
  are non-empty.
  Moreover, $\DD_2$ includes all rows
  $d$ that are continued by the rows $O_j$ with $j>i_1$ of $C(\s_0)$.

       Now, let us divide $\DD_2$ into two non-empty subsets
    $\DD_{2,1}$, $\DD_{2,2}$ according to the value taken by $\Lambda(2,d)$.
We define $\DD_{2,1}\equiv\{d\in \DD_2:\Lambda(2,d)\neq 0\}$, and
     $\DD_{2,2}\equiv\{d\in \DD_2:\Lambda(2,d)=0\}$.
Note that the row $d^{i_2}$ is an element of $\DD_2$ by the observation
    made above, while using \eqv(abn) with $s=2$,
we get, as before that $\Lambda(2,d^{i_2})=-1$, and for all $j>i_2$,
 $\Lambda(2,d^{j})=0$.
 Thus $\DD_{2,1}$ and $\DD_{2,2}$ are
    non-empty.
   In addition to that, for any row  $d$
    continued by $O_j$ with $j>i_2$ we have again
      by \eqv(abn) with $s=2$
     $\Lambda(2,d^j)=0$.
    Hence, $\DD_{2,1}$ and $\DD_{2,2}$ are non-empty,
   and $\DD_{2,2}$ contains all rows $d$
   continued by $O_j$ with $j>i_2$ of $C(\s_0)$.

       Using \eqv(abn) for $s=3$ we can again split
   $\DD_{2,2}$  into two non-empty subsets
    $\DD_{2,2,1}$ with $\Lambda(d,3)\ne 0$ and
    $\DD_{2,2,2}$ with $\Lambda(d,3)=0$.
     Furthermore, $\DD_{2,2,2}$ contains all rows that
  are continued by $O_j$ with $j>i_3$ of $C(\s_0)$, etc.
The same procedure can be repeated up
to the step  $s=k-1-t-1$. In this way we have subdivided
$\DD_2$ into $k-1-t-1$ disjoint non-empty subsets. Together with
$\DD_0$ and $\DD_1$, these constitute $k-t$ disjoint subsets $\DD_i$.
This
   proves the assertion (i).

  Let us now take any $\s$
   such that $\hbox{\rm rank }[D_N \cup C(\s)]=r+t$
    and with $\a(\s)=\a_0$.
   Assume that   $d$ and $\tilde d$ are continued by the same
   row $O_j$ of $C(\s)$ in $D_N\cup C(\s)$.
 Since $d$ and $\tilde d$  belong to different subsets
   $\DD_i$,
   for some $u \in \{1,\ldots, k-t-1\}$,
$\Lambda(d, u) \ne \Lambda(\tilde d, u)$.
    Then, writing \eqv(abn) with $s=u$
  along the row $d$ continued by $O_j$ and
  along the row $\tilde d$ continued by $O_j$
  we would get either the system
   $$
\eqalign{
\Lambda(d,u)&=0\cr
\Lambda(\tilde d,u)&=0
}
$$
 if $j>i_u$,
  or
$$
\eqalign{
\Lambda(d,u)&=-1\cr
 \Lambda(\tilde d,u)&=-1
}
$$
if $j=i_u$, or
$$
\eqalign{
\Lambda(d,u)-a_{q+j}(u)&=1\cr
 \Lambda(\tilde d,u)-a_{q+j}(u)&=1
}
$$
   if $j=i_u-1$,
 or finally
     $$
\eqalign{
\Lambda(d,u)-a_{q+j}(u)+a_{q+j+1}(u)&=0
\cr
 \Lambda(\tilde
   d,u)-a_{q+j}(u)+a_{q+j+1}(u)&=0,
}
$$
  if $j<i_u-1$. But no one of these four systems has a solution
 if $\Lambda(d, u) \ne \Lambda(\tilde d, u)$.
This proves (ii). \endproof

  By (ii) of Lemma \thv(AB.1), for any $\s$ such that
   $\hbox{\rm rank }[D_N \cup C(\s)]=r+t$ and  $\a(\s)=\a_0$
      the set of  rows of the matrix $D_N$
  is divided into $m\geq k-t$ non-empty disjoint
  subsets $\DD_1,\ldots, \DD_m$
  and the set of $k$ rows of the matrix $C(\s)$
  is divided into $m$ non-empty disjoint
   subsets $\CC_1,\dots,\CC_m$  of cardinalities
$s_1,\ldots, s_m \geq 1$, respectively,
  such that the  rows in $\CC_j$ continue the rows of $\DD_j$ only.
      But $s_j$ rows of the matrix $C(\s)$ must  be present
   $Ns_j/k$ times. Thus, first of all, in the matrix $D_N$,
  these $r_j$ rows
  must  be present $Ns_j/k$ times as well, for all
   $j=1,\ldots,m$. Thus, the number of configurations $\s$
  with  $\hbox{\rm rank }[D_N \cup C(\s)]=r+t$ such that    $\a(\s)=\a_0$
    does not exceed $\prod_{j=1}^{m} {Ns_{j}/k \choose N/k}
    {N(s_j-1)/k \choose N/k}\cdots {N/k \choose N/k}
      =( (N/k)!)^{-k} \prod_{j=1}^{m}(Ns_{j}/k)!$ which
      is bounded by
         $( (N/k)!)^{-k}  ((N(k-m+1)/k)!) ((N/k)!)^{m-1}$
          for any
  $s_1,\ldots, s_m \geq 1$ with $s_1+\cdots +s_m=k$.

          By (i) of Lemma \thv(AB.1) we have  $k-t\leq m\leq k$, so
          that
     $$
 \frac{  ((N(k-m+1)/k)!) ((N/k)!)^{m-1}}
 {((N/k)!)^{k}}
 =
     {N(k-m+1)/k \choose N/k}{N (k-m)/k \choose N/k}
    \cdots {N/k \choose N/k}$$
$$ \leq {N(t+1)/k \choose N/k}{N t/k \choose N/k}
    \cdots {N/k \choose N/k}
     = \frac {(N(t+1)/k)! }{( (N/k)!)^{(t+1)} }.$$
    Hence, for any matrix $D_N$ composed of $R$ different columns
 $$
\#\{\s: \hbox{\rm rank }[D_N\cup C(\s)]=r+t,\ \a(\s)=\a_0\}
$$
$$
\leq \Big(\sum\limits_{m=k-t}^{k}
  \sum\limits_{r_1,\ldots, r_m\geq 1,
    \atop r_1+\cdots+ r_m=R}
\sum\limits_{s_1,\ldots,s_m\geq 1,\atop
     s_1+\cdots +s_m =k}\Big)  {N(t+1)/k \choose N/k}{N t/k \choose N/k}
    \cdots {N/k \choose N/k}
=C \frac{ (N(t+1)/k)!}{ ((N/k)!)^{t+1}}.
\Eq(ik.5)
$$
  \endproof

\noindent{\it Proof of Lemma~\thv(pr4) }
     The statement \eqv(p3) is an immediate consequence of \eqv(p2)
   and \eqv(df), \eqv(5a)  if $\delta>0$ is small enough.

       The proof of \eqv(p1) and \eqv(p2) mimics the standard
    proof of the Berry-Essen inequality.
 Namely, we use the representation
    \eqv(rtt) of $ f_N^{\s_1,\ldots, \s^l}(\{t_i^j\})$
       as a product of $N$
   characteristic functions where at most $k^l$ of them
  are different. Each of them
    by standard Taylor expansion
$$  f_{N,n}^{\s^1,\ldots, \s^l}(\{t_{\b,j}\})=1-
  \frac{\Big( \sum\limits_{i=1,\ldots, k-1 \atop
   j=1, \ldots,l }(\1_{\{\s_n^j=i\}}
    -1_{\{\s_n^j=i+1\}})t_{\b,j}\Big)^2
}{4 (N/k) \var X }\var  X$$
$$
-\theta_n
     \frac{i\Big( \sum\limits_{i=1,\ldots, k-1 \atop
   j=1, \ldots,l }(\1_{\{\s_n^j=i\}}
    -1_{\{\s_n^j=i+1\}})t_{\b,j}\Big)^3
}{6 ((2N/k) \var X)^{3/2} }\E(X-\E X)^3\equiv 1-r_n
\Eq(p11)
 $$
  with $|\theta_n|<1$.
  It follows that $|r_n|<C_1\|\vec t\|^2 N^{-1}+
     C_2 \|\vec t\|^3 N^{-3/2}$, for some $C_1, C_2>0$, all
     $\s^1, \ldots, \s^l$, and all $n$.
    Then $|r_n|<1/2$ and
    $|r_n|^2 < C_3 \|\vec t\|^3 N^{-3/2}$, for
    some $C_3>0$ and  all $\vec t$  satisfying
  $\|\vec t\|<\delta \sqrt{N}$, with  $\delta$ enough small.
   Thus, $\ln  f_{N,n}^{\s^1,\ldots, \s^l}(\{t_{\b,j}\})=
      -r_n+ \tilde \theta_n r_n^2/2$
    (using the expansion $\ln (1+z)=z+\tilde \theta z^2/2$
     for $\|z\|<1/2$ with $\|\tilde \theta\|<1$) ,  with some
   $|\tilde \theta_n|<1$ for all $\s^1, \ldots, \s^l$,
       all $n$, and all $t$ satisfying $\|\vec t\|<\delta \sqrt{N}$.
    It follows that
$ f_{N}^{\s^1,\ldots, \s^l}(\{t_{\b,j}\})
   =\exp(-\sum_{n=1}^{N}r_n+\sum_{n=1}^{N}
     \tilde \theta_n r_n^2/2)$.
  Here $-\sum_{n=1}^{N}r_n= -\vec t B_N(\s^1, \ldots, \s^l) \vec t/2
         +\sum_{n=1}^N p_n$ where $|p_n| \leq C_2
     \|\vec t\|^{3} N^{-3/2}$.
  Then
$$
f_{N}^{\s^1,\ldots, \s^l}(\{t_{\b,j}\})
   =e^{  -\vec t B_N(\s^1, \ldots, \s^l) \vec
  t/2} e^{\sum_{n=1}^N (p_n+\tilde\theta_n r_n^2/2)},
\Eq(p12)
$$
    where $|p_n|+|\tilde \theta_n r_n^2/2|\leq (C_2 +C_3/2)\|\vec t\|^3
  N^{-3/2}$.
     Hence
  $|e^{\sum_{n=1}^N (p_n+ \tilde \theta_n r_n^2/2)}-1|\leq C_4 \|\vec t\|^3
  N^{-1/2}$, for all $\vec t$ satisfying $\|\vec t\|<\epsilon N^{1/6}$
   with $\epsilon>0$ small enough.
     Moreover, $|\sum_{n=1}^N (p_n+ \tilde \theta_n r_n^2/2)|\leq C_5
        \|\vec t\|^3 N^{-1/2}$, which implies   \eqv(p2).
 This concludes the proof of Lemma \thv(pr4)
\endproof

\bigskip

\chap{5. The unrestricted partioning problem.}5

     In the previous section we considered the state
  space of spin configurations where the number of spins
  taking each of $k$ values is exactly $N/k$.
      Here we want to discuss what happens if all partitions are
  permitted.
Naturally, we divide again the space of all configurations
  $\{1,\dots,k\}^N$  into equivalence classes
  obtained by permutations of spins.
   Thus our state space $\tilde \Sigma_N$  has $k^N (k!)^{-1}$ elements.
       Let us  define the random variables
         $ Y^\b(\s)$ as in the previous section, see  \eqv(az).
   Then we may state the following conjecture analogous to
        Theorem~\thv(1.1th).

\noindent{\bf Conjecture \TH(C.1) } {\it Let
$$
\tilde V^{\b}(\s) =k^{N/(k-1)}N^{-1/2}k^{1/2}(k!)^{-1/(k-1)}
    \pi^{-1/2}\sqrt{3} | Y^{\b}(\s)|, \ \ \ \ \b=1,\ldots, k-1.
\Eq(v.3)
$$
  Then the point process on $\R_+^{k-1}$
$$
\sum_{\s \in \tilde \S_N}
    \delta_{(\tilde V^{1} (\s),\ldots, \tilde V^{k-1}(\s))}
$$
converges to the Poisson point process on $\R_+^{k-1}$
    with the intensity measure which is the Lebesgue measure.
}

  Using Theorem~\thv(mainth), the assertion of the conjecture would be
  an immediate consequence of the following conjecture, that is the
  analogue of  Proposition~\thv(mainl).

\noindent {\bf Conjecture \TH(C.2)} {\it   Denote by $\sum_{\s^1,\ldots,\s^l \in \S_N}(\cdot)$ the sum
  over all possible ordered sequences of {\sl  different} elements
   of $\S_N$.   Then for any $l=1,2,\ldots,$
  any constants $c_j^{\b}> 0$, $j=1,\ldots, l$,
   $\b=1,\ldots, k-1$ we have:
 $$
 \sum_{\s^1,\ldots, \s^l \in \S_N   }
  \P\Big( \forall \b=1,\ldots, k-1, \forall j=1,\ldots, l
      \frac{ | Y^{\b} (\s^j)| }{ \sqrt{ 2(N/k) \hbox{\rm var}\,X } }
     < \frac{ c_j^{\beta} } {(k!)^{-1/(k-1)} k^{N/(k-1)}} \Big)
$$
$$
 \to  \prod\limits_{j=1,\ldots,l \atop \b=1,\ldots, k-1}
          \frac{ 2  c_j^\b  \sqrt{ \hbox{\rm var}\, X   } } {
             \sqrt{ 2\pi \E (X^2)}  }.
\Eq(jkh)
$$
}

\remark
 One can notice the difference between  the right-hand sides
   of \eqv(jk) and \eqv(jkh). In spite of this difference,
the proof of this statement proceeds along the same
  lines as that of Proposition~\thv(mainl).
        The only point that we were not able to complete
    is that the sum analogous to
  $J_N^2$ in \eqv(dva1)
(recall that it is a sum over sets $\s^1,\ldots, \s^l$
    such that the system  $\{ Y^\b(\s^j)\}_{j=1,
   \ldots,l,\atop \b=1,\ldots k-1}$ is linearly dependent
    and, moreover, for any basis of this system there
   exists a number $j\in \{1,\ldots, l\}$ such that
      for  some
     non-empty subset of coordinates $\b\in \{1,\ldots,k-1\}$
    the random variables
      $ Y^\b(\s^j)$ are included in this basis and for
    some  non-empty subset of coordinates $\b\in \{1,\ldots,k-1\}$
     they are not included there)
 converges to $0$  as $N\to \infty$.
  Therefore the whole statement remains a conjecture.

 \remark {\bf The case $k=2$.}
 In the case $k=2$ the sum $J_N^2$ is absent.
  Hence, in this case we can provide an entire
  proof of \eqv(jkh) and therefore prove
  our conjecture.  The result in the case $k=2$ is not new: it
    has been already established by Ch.~Borgs, J.~Chayes and
   B.~Pittel in \cite{BCP}, Theorem~2.8.
    Our Theorem~\thv(mainth) gives an alternative proof
    for it via \eqv(jkh).

       Finally we  sketch the arguments that should lead to  \eqv(jkh)
  and explain the differences with \eqv(jk).
  To start with, similarly to \eqv(dva),  we split
 $$
\sum_{\s^1,\ldots, \s^l \in \tilde \S_N \atop
     \hbox{\rm rank}C(\s^1, \ldots, \s^l)=(k-1)l}\P(\cdot)
    +\sum_{\s^1,\ldots, \s^l \in \tilde \S_N \atop
     \hbox{\rm rank}C(\s^1, \ldots, \s^l)<(k-1)l}\P(\cdot).
   \Eq(hm)
$$
   We are able to prove  that the first part of \eqv(hm)
      converges to the left-hand side of \eqv(jkh).
   For that purpose, we introduce again ``the main part''
    of the state space with $\a\in(0, 1/2)$:
$$
\tilde \RR^ {\a}_{l,N}= \Big\{\s^1, \ldots, \s^l \in \Sigma_N:
       \forall 1\leq j \leq l,  \forall 1\leq i<r\leq l,
     1\leq \b,\gamma,\eta \leq k, \beta\ne \g$$
$$ \Big|\sum_{n=1}^N \1_{\s_n^j=\b}-N/k\Big|< N^{\a} \sqrt{N},
  \Big|\sum_{n=1}^N (\1_{\{\s_n^i=\b\}}-\1_{\{\s_n^i=\gamma\}})
    \1_{\{\s_n^r=\eta\}}\Big|<N^{\a}\sqrt{N}
\Big\}
\Eq (8.2)
$$
   where
 $$
 \|\tilde \RR^{\a}_{l,N}\|\geq k^{Nl}(1-\exp(-h N^{2\a}))(k!)^{-l}
\Eq(45)
$$
and  split the first term of \eqv(hm) into two terms
  $$
\sum_{\s^1,\ldots, \s^l \in \tilde \RR^\a_{l,N} }\P(\cdot)
    +\sum_{\s^1,\ldots, \s^l \not\in \tilde \RR^\a_{l,N}\atop
     \hbox{\rm rank}C(\s^1, \ldots, \s^l)=(k-1)l}\P(\cdot).
\Eq(hm1)
$$
   The  second term of \eqv(hm1) converges to zero exponentially fast:
  the number of configurations in it
      is at most  $O(\exp(-h N^{2\a})k^{Nl})$ by \eqv(45),
  while the probability  $\P(\cdot)=O( N^{l} k^{-Nl})$ by
      the analogue of Lemma~\eqv(pr1).

      To treat the first term  of \eqv(hm), let us stress
     that an important difference compared to the previous
     sections
is the fact  that the variables
  $ Y^{\b}(\s)$  are now not necessarily centered. Namely,
$$
 \E  Y^{\b}(\s)=(\E X) \sum_{n=1}^N  (\1_{\{\s_n=\b\}}-
\1_{\{\s_n=\b+1\}})=\E X \left[\#\{n: \s_n=\b\}- \#\{n: \s_n=\b+1\}
\right]
 \Eq(jkj)
$$
as    it may
   happen that $\#\{n: \s_n=\b\}\ne \#\{n: \s_n=\b+1\}$.

 Taking this observation into account  and proceeding similarly to  the analysis
  of \eqv(55), we can  show
   that, uniformly for all
 $  \s^1,\ldots, \s^l\in
     \tilde \RR^\a_{l,N}$,
  $$
\P(\cdot)=\frac{ k^{-Nl} (k!)^l
 \prod_{j,\b} (2 c_j^\b) }{(2\pi)^{(k-1)l/2} }
      \exp\Big(- \frac{ \E \vec  Y_j^\b}{\sqrt{2 (N/k) \var  X }} \frac{B^{-1}
       }{2} \frac{ \E\vec  Y_j^\b}{\sqrt{2(N/k) \var  X
      }}\Big )+o(k^{-Nl})
\Eq(19)
$$
       where the matrix $B$ consists of $l$ diagonal blocks
   $(k-1)\times (k-1)$, each block  having $1$ on the diagonal,
   $-1/2$ on the line under the diagonal and $0$ everywhere else.
   Thus the first term of \eqv(hm1) by \eqv(19) and  \eqv(45)  equals
$$
\eqalign{
& \sum_{\s^1, \ldots, \s^l \in \tilde \RR^\a_{l, N}}
      \frac{ k^{-Nl} (k!)^l
     \prod_{j,\b} (2 c_j^\b) }{(2\pi)^{(k-1)l/2} }
      \exp\Big(- \frac{ \E \vec  Y_j^\b}{\sqrt{2(N/k) \var  X
}} \frac{B^{-1}
       }{2} \frac{ \E\vec  Y_j^\b}{\sqrt{2(N/k) \var  X
      }}\Big )+o(1)
\cr&  =
        \frac{\prod_{j,\b} (2 c_j^\b) }{(2\pi)^{(k-1)l/2} }
     E_{\s^1, \ldots, \s^l  }
      \exp\Big(- \frac{ \E \vec  Y_j^\b}{\sqrt{2(N/k) \var  X }} \frac{B^{-1}
       }{2} \frac{ \E\vec  Y_j^\b}{\sqrt{2(N/k) \var  X
      }}\Big )+o(1).
}
\Eq(mq1)
$$
 By the Central Limit Theorem  the vector
    $ \sum_{n=1}^{N}(\1_{\{\s_n^j=\b\}}-
     \1_{\{\s_n^j=\b+1\}})/
      \sqrt{2 N/k}$ on
   $\tilde \Sigma_N^{\otimes l}$
  converges to a Gaussian vector $Z_j^\b$  with zero mean
     and covariance matrix $B$ as $N \to \infty$. Hence, \eqv(mq1)
   converges to
$$
\frac{\prod_{j,\b} (2 c_j^\b) }{(2\pi)^{(k-1)l/2} }
     \E_Z \exp\Big( -\frac{\E X}{\sqrt{ \hbox{\rm var}\, X} }
       \vec Z_j^\b \frac{B^{-1}}{2}\vec Z_j^\b
  \frac{\E X}{\sqrt{\hbox{\rm var}\, X} }
\Big) =
     \prod_{j,\b} \frac{ 2 c_j^\b
     \sqrt {\hbox{\rm var}X  } }{\sqrt{ 2\pi} \sqrt{(\E X)^2+ \hbox{\rm var}\,X
 }    }
\Eq(yh)
$$
  which is the right-hand side of \eqv(jkh).
     This finishes the analysis of the first term of \eqv(hm).

      To treat the second term, we  split it into two parts
   $J_N^1$ and $J_N^2$ analogously to \eqv(dva1).
       The analysis of $J_N^1$ is exactly the same
  as in the proof of Proposition~\thv(mainl) and relies on
       Lemmatas~\eqv(pr1) and \eqv(pr2).

       However, the problem  with the sum $J_N^2$ persists.
    First of all, this sum  contains much more
   terms than in the case of the previous section  as
   it consists essentially of configurations $\s^1,\ldots, \s^l$
   where some of the values of spins $\b$  among $\{1,\ldots, k\}$
   figure out more often than others, i.e.
    $\#\{n: \s_n=\b\}> \#\{n: \s_n=\b+1\}$.
    Lemma \thv(pr3)  is  not valid anymore. Second,
   for all such configurations $\s$,
   the random variables
   $ Y^{\b}(\s)$ are not centered and
   consequently the estimate of the probability
    $\P(\cdot)$  suggested by Lemma~\thv(pr1)
   is too rough. We
   did not manage to complete the details of this analysis.

\bigskip

\chap{References.} 0

\frenchspacing
\item{[BFM]} H.~Bauke, S.~Franz, and St.~Mertens,
 Number partitioning as random energy model, {\it Journal of
 Stat. Mech.: Theor. and Exp.}, P04003 (2004).
\item{[BBG1]}  G.~Ben Arous, A.~Bovier, V.~Gayrard.
  Glauber Dynamics for the Random Energy Model~1.
   Metastable motion on extreme states. {\it Commun. Math. Phys.}
  {\bf 235}, 379--425 (2003).
\item{[BBG2]} G.~Ben Arous, A.~Bovier, V.~Gayrard.
   Glauber Dynamics for the Random Energy Model~2. Aging
  below the critical temperature. {\it Commun. Math. Phys.}
   {\bf 235 }  (2003).
\item{BCMP} C. Borgs, J.T. Chayes, S. Mertens, and B. Pittel. 
Phase diagram for the constrained integer partitioning problem, {\it   
Random Structures Algorithms}  {\bf 24}, 315--380  (2004). 
\item{[BCP]} Ch.~Borgs, J.T.~Chayes, and B.~Pittel,
Phase transition and finite-size scaling for the integer partitioning
problem. Analysis of algorithms (Krynica Morska, 2000).
{\it Random Structures Algorithms} {\bf 19}, 247--288 (2001).
\item{[BM]} A.~Bovier and D.~Mason,
 Extreme value behavior in the Hopfield model, {\it Ann. Appl. Probab.}
{\bf   11}, 91--120 (2001).
\item{[Fe]} W.~Feller, An introduction to probability theory and its
applications 1, Wiley, New York, 1957.
\item{[Gal]} J.~Galambos, On the distribution of the maximum of random
variables, {\it Ann. Math. Statist.} {\bf 43}, 516-521 (1972).
\item{[JK]} I.~Junier and J.~ Kurchan, Microscopic realizations of the trap
model. {\it J. Phys. A} {\bf 37}, 3945--3965 (2003).
\item{[Kal]} O.~Kallenberg, Random measures, Akademie Verlag, Berlin, 1983.
\item{[KP]}  H. Koch and J. Piasko, ``Some rigorous results on the Hopfield
neural network model'', J. Stat. Phys. {\bf 55},  903-928 (1989).
\item{[LLR]} M.R.~Leadbetter, G.~Lindgren, and H.~Rootz\'en,
Extremes and related properties of random sequences and processes,
Springer, Berlin, 1983.
\item{[Mer1]} St.~Mertens,
Random costs in combinatorial optimization, {\it Phys. Rev. Letts.}
{\bf 84}, 1347-1350 (2000).
\item{[Mer2]} St.~Mertens,
A physicist's approach to number partitioning.
Phase transitions in combinatorial problems (Trieste, 1999),
{\it
Theoret. Comput. Sci.} {\bf  265}, 79--108 (2001).

\end